%% file: cas-sc-template.tex
\documentclass[a4paper,fleqn]{cas-sc}
\usepackage[numbers,sort&compress]{natbib}
\usepackage[utf8]{inputenc}
\usepackage{url}
\usepackage{nicefrac}
\usepackage{microtype}
\usepackage{float}
\input{our_imports}

\begin{document}
\let\WriteBookmarks\relax
\def\floatpagepagefraction{1}
 \def\textpagefraction{.001}

\shorttitle{DIPA for Solving Imaging Inverse Problems}
\shortauthors{R. Gualdr\'on-Hurtado et~al.}

\title[mode=title]{DIPA: Distilled Preconditioned Algorithms for Solving Imaging Inverse Problems}
\tnotetext[1]{The conference precursor of this work was presented at the 2025 IEEE International Workshop on Computational Advances in Multi-Sensor Adaptive Processing (CAMSAP)~\cite{gualdron2025deep}.}

\author[1]{Romario Gualdr\'on-Hurtado}

\author[2]{Roman Jacome}

\author[1]{Leon Suarez}

\author[1]{Henry Arguello}[orcid=0000-0002-2202-253X]

\affiliation[1]{organization={Department of Computer Science, Universidad Industrial de Santander},
            city={Bucaramanga},
            postcode={680002},
            country={Colombia}}

\affiliation[2]{organization={Department of Electrical Engineering, Universidad Industrial de Santander},
            city={Bucaramanga},
            postcode={680002},
            country={Colombia}}

\cortext[cor1]{Corresponding author henarfu@uis.edu.co}

\begin{abstract}
Solving imaging inverse problems has usually been addressed by designing proper prior models of the underlying signal. However, minimizing the data fidelity term poses significant challenges due to the ill-conditioned sensing matrix caused by physical constraints in the acquisition system. Thus, preconditioning techniques have been adopted in classical optimization theory to address ill-conditioned data-fidelity minimization by transforming the algorithm gradient step to achieve faster convergence and better numerical stability. {We extend the preconditioning concept beyond convergence acceleration and use it to improve reconstruction quality. We introduce \textit{DIPA: Distilled Preconditioned Algorithms}, where a preconditioning operator (PO) is optimized using teacher-guided distillation criteria. Unlike standard model-compression KD, the teacher and student differ by the sensing operators available during reconstruction: the teacher uses a simulated, better-conditioned, and more informative sensing matrix, whereas the student uses the physically feasible sensing matrix.} We design different distillation loss functions to transfer different properties of the teacher algorithm to the preconditioned student. The PO can be linear (L-DIPA), allowing interpretability, or non-linear (N-DIPA), parametrized by a neural network, offering better scalability. We validate the proposed PO design across several imaging modalities, including magnetic resonance imaging, compressed sensing, and super-resolution imaging.
\end{abstract}



\maketitle

\section{Introduction}
Inverse problems involve reconstructing an unknown signal \( \mathbf{x}  \in \mathbb{R}^{n} \)  from noisy, corrupted, or usually undersampled observations  \( \mathbf{y} \in \mathbb{R}^m (m\ll n)\) , thus, the recovery process is generally non-invertible and ill-posed. In this work, we focus on linear inverse problems where the forward model is modeled by a sensing matrix \( \mathbf{A} \in \mathbb{R}^{m \times n} \), such that \( \mathbf{y} = \mathbf{A} \mathbf{x} + \mathbf{e} \), where \( \mathbf{e} \) is additive noise {\cite{bertero1985linear}}. Numerous imaging tasks are based on these principles, including image restoration such as deblurring, denoising, inpainting, or superresolution \cite{gunturk2018image}, compressed sensing \cite{zha2023learning, cs}, medical imaging applications such as magnetic resonance imaging (MRI) \cite{lustig2008compressed}, and more (see \cite{IPDL,bai2020deep,bertero2021introduction} and references therein). {The reconstruction of \( \mathbf{x} \) is typically formulated as 

\begin{equation}
    \mathbf{\hat{x}}=\argmin_{\mathbf{x}} g(\mathbf{x}) + \lambda h(\mathbf{x}),\label{eq:optmization_problem}
\end{equation} 

where $g(\mathbf{x})$ is the data-fidelity term, usually $g(\mathbf{x})=\frac{1}{2}\|\mathbf{A}\mathbf{x}-\mathbf{y}\|_2^2$, and $h(\mathbf{x})$ is a regularization function that incorporates prior knowledge about \( \mathbf{x} \), such as sparsity, smoothness, or low-rankness. The scalar parameter \( \lambda > 0 \) balances data consistency and prior regularization. Common choices for $h(\mathbf{x})$ include Tikhonov regularization \cite{golub1999tikhonov}, sparsity-promoting regularization \cite{jin2017sparsity}, and total variation \cite{strong2003edge}.}

\begin{figure}[pos=!tbp]
\centering
     \includegraphics[width=0.7\textwidth]{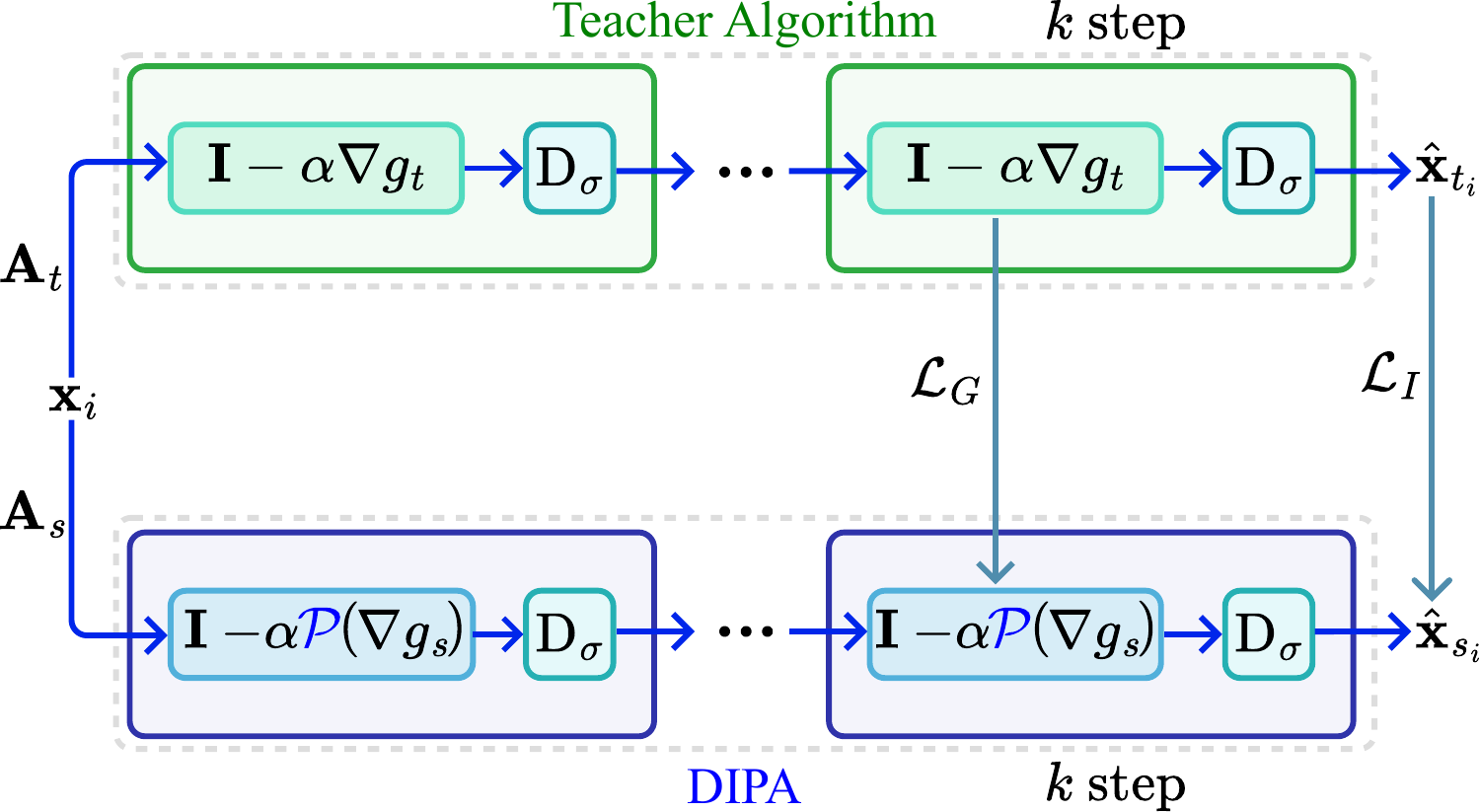}

    \caption{{DIPA framework. A teacher algorithm uses a simulated, better-conditioned sensing matrix and transfers reconstruction behavior through output imitation and data-fidelity gradient alignment. The deployable student uses the physically feasible sensing matrix and a learned preconditioning operator (PO).}}
    \label{fig:MAIN_FIG}

    \end{figure}

     While classical variational methods rely on explicit regularizers, the plug-and-play (PnP) framework integrates model-based recovery methods with precise forward modeling of the physical acquisition phenomenon with a wide range of data priors. PnP traces back its roots to proximal algorithms \cite{proximal}, where these operators, usually defined by analytical models of the underlying signals such as sparsity or low-rank {\cite{zha2023learning}}, are replaced by a general-purpose image denoiser operator {\cite{teodoro2019image}}. This approach allows the integration of classical image denoiser operators {\cite{ma2023improving} such as BM3D \cite{BM3D}, NLM \cite{bcm_nlm}, RF \cite{gastal2011domain}} and current deep learning (DL) denoisers {\cite{zhang2021plug}}.  Additionally, PnP can adopt different optimization solvers to minimize the data-fidelity term and the implicit regularization function (defined by the selected image denoiser), such as ADMM {\cite{boyd2011distributed}}, HQS {\cite{he2013half}}, FISTA \cite{FISTA}, PGD {\cite{daubechies2004iterative}}, among others.While most advancements in PnP approaches have focused on high-performance denoisers as implicit image priors, the data-fidelity step remains a bottleneck. This step is affected both by ill-conditioning, which slows and destabilizes iterative optimization, and by ill-posedness or rank deficiency, which creates non-unique solutions in the null space of the sensing operator. These issues arise in acquisition systems with severe physical limitations, such as high acceleration factors in MRI, high compression rates in compressive imaging, or strong downsampling in super-resolution.

Preconditioning techniques can be divided into several types, each suited to different applications. Linear preconditioning, based on the transformation of the system using linear operators, preconditioning matrices (PMs), includes methods such as Incomplete Cholesky and Jacobi preconditioners \cite{chen2021multiscale,anzt2019adaptive,garber2024image}. Nonlinear preconditioning introduces a nonlinear operator that dynamically adjusts based on the specific characteristics of the problem, adapting to the data or iteration process to enhance performance \cite{gander2017origins,doi:10.1137/17M1128502}. Gradient preconditioning modifies the gradient in optimization algorithms, significantly improving convergence speed and stability, making it particularly useful in complex or high-dimensional optimization tasks \cite{andrea2016preconditioning,fessler1999conjugate}. Various design approaches have been explored in the state-of-the-art to solve inverse problems. Krylov subspace methods approximate the sensing matrix to construct the PM, improving system conditioning \cite{garber2024image,chan1984nonlinearly,pearson2020preconditioners}. The PM can take various forms, such as a regularized inverse formulation, which improves numerical stability but remains sensitive to parameter choices that may limit robustness across different problem settings. Polynomial preconditioning \cite{iyer2024polynomial} builds the PM using a polynomial function to equalize the eigenvalue distribution of the \textit{gram} matrix of the sensing operator, resulting in a transformation of the system that is not explicitly data-driven and may suffer from reduced performance. In contrast, learned preconditioning methods leverage a dataset to approximate the PM, allowing it to adapt to the specific characteristics of both the data and the inverse problem \cite{ehrhardt2024learning}. While these methods aim to accelerate convergence, {they do not by themselves resolve the ambiguity caused by the sensing null space \cite{gualdron2026gsnr,jacome2026npn}; the final reconstruction still depends on how the data-fidelity direction interacts with the image prior.}

{To extend PO design beyond convergence acceleration, we introduce \textit{DIPA: Distilled Preconditioned Algorithms for Solving Imaging Inverse Problems}, which uses a simulated, better-conditioned teacher sensing matrix to guide the preconditioner used with a physically feasible student sensing matrix.}

{Knowledge distillation (KD) is commonly used when a strong teacher model guides a constrained student model \cite{hinton2015distillingknowledgeneuralnetwork,gou2021knowledge}. In this paper, ``distillation'' means teacher-guided behavior matching rather than neural-network compression: a virtual sensing matrix with fewer constraints is used only during training to produce teacher reconstructions and teacher data-fidelity gradients. The learned PO is then deployed with the constrained sensing matrix used by the student acquisition system.} 

{Our goal is to design a PO $\mathcal{P}$ that makes the student reconstruction behave closer to the teacher reconstruction under the image distribution and reconstruction prior used during training. DIPA learns a data- and prior-dependent gradient transformation for the deployable operator.} To achieve this, we developed different distillation loss functions based on the teacher reconstruction, data-fidelity gradients, and algorithm convergence criteria for  PO optimization.  We devised two designs for the PO; the first one is a linear operator (preconditioning matrix), which provides high interpretability, but it is not scalable for large-scale inverse problems. We denoted this approach L-DIPA. In the second, the PO is parameterized by a non-linear neural network, allowing for well-scalability to increases in the signal dimensionality but hindering interpretability and analysis. We denote this formulation N-DIPA. The proposed method was validated in several computational imaging scenarios, such as compressed sensing (CS), MRI, and super-resolution (SR), significantly improving classical preconditioning and no-preconditioned approaches. Fig.~\ref{fig:MAIN_FIG} summarizes the proposed teacher-student framework. {To the best of our knowledge, this is the first work that uses teacher-guided distillation to learn a preconditioner for PnP/RED reconstruction algorithms.} Preliminary results for this method were presented in \cite{gualdron2025deep}. 
\noindent Our contributions are the following:

\begin{enumerate}
    \item {A teacher-guided bilevel framework that transfers the reconstruction behavior of a PnP/RED algorithm using a virtual, high-performance sensing matrix to a PO used with a physically implementable, constrained sensing matrix.}
    \item Different distillation loss functions are proposed to match the student's and teacher's outputs and data fidelity gradients. 
    \item {Linear and nonlinear preconditioning schemes are formulated, trading interpretability and scalability depending on the application.}
    \item {A convergence-motivated regularization term is introduced for the linear PO to encourage student updates that remain close to the teacher updates.}
\end{enumerate}
\section{Background}
\label{sec:background}

\subsection{Plug-and-Play algorithms}

PnP algorithms provide a flexible framework for solving computational imaging
inverse problems by combining a model-based data-fidelity step with an image
denoising operator that acts as an implicit prior \cite{venkatakrishnan2013plug}.
They are rooted in proximal algorithms \cite{proximal}, which are commonly used to solve composite optimization problems of the form in \eqref{eq:optmization_problem}, where $g(\mathbf{x})$ is smooth and $h(\mathbf{x})$ may be non-smooth. In this setting, the data-fidelity term is
typically handled through a gradient or least-squares update, while the regularization term is handled through its proximal operator. For example,
proximal-gradient methods such as FISTA \cite{FISTA}, and variable-splitting methods such as ADMM \cite{boyd2011distributed} and HQS \cite{he2013half}, separate the influence of $g(\mathbf{x})$ and $h(\mathbf{x})$ so that the non-smooth regularizer can be treated through a
proximal step.

The key idea of PnP is to replace this proximal step with a general-purpose denoiser $\mathrm{D}_\sigma$. This substitution enables the use of powerful handcrafted denoisers, such as BM3D \cite{BM3D}, or neural-network-based denoisers \cite{burger2012image}, as implicit image priors even when they do not correspond to an explicit regularization function.In the case that $g(\mathbf{x})=\|\mathbf{y} -\mathbf{A}\mathbf{x}\|_2^2$, with $\alpha>0$ as the stepsize, the proximal step is replaced by
\begin{equation}
    \label{eq:denoising_step}
    \mathbf{x}^{k}=\mathrm{D}_\sigma(\mathbf{x}^{k-1} - \alpha \mathbf{A}^\top (\mathbf{A}\mathbf{x}^{k-1} - \mathbf{y})).
\end{equation}

An alternative approach, Regularization by Denoising (RED) {\cite{RED}}, also leverages {{image}} denoisers to address imaging inverse problems. RED introduces a smoothness-based regularization term $h(\mathbf{x})=\frac{1}{2}\mathbf{x}^\top (\mathbf{x} - \mathrm{D}_\sigma(\mathbf{x}))$, defined by the inner product of the image and its denoising residual. The gradient of $h(\mathbf{x})$ is approximated as the residual $\nabla h(\mathbf{x}) = \mathbf{x} - \mathrm{D}_\sigma(\mathbf{x})$. In contrast to PnP methods, which are limited in their choice of optimization techniques, RED offers more freedom in selecting an optimization method for solving the inverse problem. For instance, using a gradient descent step, the update for the RED objective is given by
\begin{equation}
    \label{eq:red_step}
    \mathbf{x}^{k}=\mathbf{x}^{k-1} - \alpha ( \mathbf{A}^\top (\mathbf{A}\mathbf{x}^{k-1} - \mathbf{y}) + \lambda (\mathbf{x}^{k-1} - \mathrm{D}_\sigma(\mathbf{x}^{k-1}))).
\end{equation}

In this work, we focus on the PnP-FISTA and RED-FISTA approaches.















\subsection{Preconditioning methods}

{Preconditioning is classically used to improve the numerical properties of iterative solvers, most commonly by reducing the effective condition number of the normal equations or by reshaping the spectrum of the operator involved in the data-fidelity step \cite{dahlke2012multilevel,nielsen2010efficient}. This is distinct from ill-posedness: rank deficiency or undersampling creates a non-trivial null space, whereas ill-conditioning concerns sensitivity and slow convergence along measured directions. In this work, the PO is therefore not presented as a replacement for regularization; it is a learned transformation of the data-fidelity update that complements the PnP/RED prior.} Preconditioning techniques can be divided into several types, each suited to different applications. {Linear preconditioning may be implemented as a left preconditioner on the residual, a right preconditioner on the unknown, or a preconditioner of the gradient/normal equation.} Nonlinear preconditioning, $\mathcal{P}_{\theta}\big({\mathbf{y}-\mathbf{A}}\mathbf{x}\big)=0$ where $\mathcal{P}_{\theta}(\cdot)$ is a preconditioning nonlinear operator, dynamically adjusts based on the specific characteristics of the problem, adapting to the data or iteration process to enhance performance \cite{gander2017origins,doi:10.1137/17M1128502}.  Gradient preconditioning (GP), with the form $\mathbf{x}^{k} =  \mathbf{x}^{k-1} - \alpha\mathcal{P}\Big(\nabla g (\mathbf{x}^{k-1} )\Big)$, modifies the gradient in optimization algorithms, significantly improving convergence speed and stability, making it particularly useful in complex or high-dimensional optimization tasks \cite{andrea2016preconditioning,fessler1999conjugate}. In this work, we focus on gradient preconditioning; the same teacher-guided principle could be adapted to other preconditioning forms. For instance, the GP of \eqref{eq:denoising_step} and \eqref{eq:red_step} is given by 
\begin{equation}
     \mathbf{x}^{k}=\mathrm{D}_\sigma(\mathbf{x}^{k-1} - \alpha \mathcal{P}(\mathbf{A}^\top (\mathbf{A}\mathbf{x}^{k-1} - \mathbf{y}))),
\end{equation}
and 
\begin{equation}
    \mathbf{x}^{k}=\mathbf{x}^{k-1} - \alpha\left(\mathcal{P}(\mathbf{A}^\top (\mathbf{A}\mathbf{x}^{k-1} - \mathbf{y})) {+ \lambda (\mathbf{x}^{k-1} - \mathrm{D}_\sigma(\mathbf{x}^{k-1}))}\right)
\end{equation}
respectively. Various design approaches have been explored to solve inverse problems in the state-of-the-art. Krylov Subspace Methods approximate the matrix $ \mathbf{A} $ to construct {a linear PO} $\mathbf{P}$, improving the {system conditioning} \cite{garber2024image,chan1984nonlinearly,pearson2020preconditioners}. {For gradient preconditioning, a dimensionally consistent linear choice is, for example, $ \mathcal{P}=(\mathbf{A}^\top\mathbf{A}+\eta\mathbf{I}_n)^{-1}$, or equivalently a left-preconditioned gradient $\mathbf{A}^\top\mathbf{M}(\mathbf{A}\mathbf{x}-\mathbf{y})$ with $\mathbf{M}\in\mathbb{R}^{m\times m}$.} Here $ \eta $ is a regularization parameter. Although this formulation improves numerical stability, its performance is susceptible to the choice of parameters, such as $ \eta $, which can limit robustness across different problem settings.

Polynomial preconditioning \cite{iyer2024polynomial} builds $ \mathcal{P} $ using a polynomial function $ p(\cdot) $, resulting in $ p(\mathbf{A}^\top \mathbf{A}) $.  This approach may suffer from reduced performance since it is not data-driven. In contrast, data-driven methods try to approximate $ \mathcal{P}_{\theta}$, enabling the PO to adapt to the specific characteristics of both the data and the inverse problem \cite{ehrhardt2024learning}. While these methods aim to accelerate convergence, minimizing the data fidelity term can introduce a non-trivial null space, which may affect the quality of the final solution and the ability to recover the original signal accurately.

\subsection{Knowledge distillation}
Knowledge distillation, popularized by \cite{hinton2015distillingknowledgeneuralnetwork}, is a neural network compression technique designed to address the challenge of deploying large deep learning models on resource-constrained devices. These devices face limitations in computational and storage requirements, which makes it impractical to deploy large deep-learning models. KD addresses this by transferring knowledge from a high-parametrized, high-performance model (the teacher) to a smaller, more efficient model (the student), whose performance drops due to parameter reduction. In KD, the teacher model guides the student model, enabling it to approximate or improve the teacher's performance despite its smaller size. KD has been mainly used for high-level computer vision tasks such as classification \cite{zagoruyko2016paying}, object detection \cite{KD_OBJECT_DETECTION}, and segmentation \cite{KD_SEGMENTATION}. However, fewer works have focused on low-level vision tasks such as image reconstruction {\cite{murugesan2020kd} or image denoising \cite{li2022multiple}.} 

The above ingredients suggest a natural but underexplored direction. PnP and RED algorithms provide flexible reconstruction frameworks, preconditioning modifies the data-fidelity descent direction, and KD provides a mechanism for transferring behavior from a stronger teacher to a constrained student.  {Unlike standard KD, the teacher and student differ not by model size, but by the sensing operator available during reconstruction; hence, DIPA should be understood as teacher-guided algorithmic distillation rather than compression of a large neural network.}

\section{Distilling knowledge of algorithms via preconditioning}

We want to design a PO $\mathcal{P}$ based on KD criteria. Before detailing the proposed optimization approach, the teacher and student models will be defined.

\subsection{DIPA Student and teacher setting}

Here, we consider that the student model of the proposed KD framework is a DIPA algorithm using a sensing matrix $\mathbf{A}_s \in \mathbb{R}^{m_s\times n}$ which produces a measurement vector $\mathbf{y}_s = \mathbf{A}_s \mathbf{x} + \mathbf{e}_s$. While the teacher model is a PnP algorithm that recovers the underlying signal $\mathbf{x}$ from a measurement vector obtained with a virtual acquisition system (infeasible to implement in practice but with high reconstruction performance) $\mathbf{A}_t \in \mathbb{R}^{m_t\times n}$ as $\mathbf{y}_t = \mathbf{A}_t \mathbf{x} + \mathbf{e}_t$. The student model requires the preconditioning operator since $\mathbf{A}_s$ is more ill-conditioned than the virtual teacher sensing matrix $\mathbf{A}_t$. {We consider three representative teacher-student sensing settings.}
\begin{example}[MRI]
Consider the single-coil MRI, where undersampled k-space measurements are acquired to reduce the acquisition time \cite{pruessmann1998coil}. Here, the teacher and the student are defined as $\mathbf{A}_s = \mathbf{M}_s\mathbf{F} \in \mathbb{C}^{n\times n}$ where $\mathbf{F}\in \mathbb{C}^{n\times n }$ is the 2D discrete Fourier transform of the signal and $\mathbf{M}_s \in \mathbb{C}^{n\times n}$ is a 2D mask sampling the k-space. Similarly, the teacher is  $\mathbf{A}_t = \mathbf{M}_t\mathbf{F} \in \mathbb{C}^{n\times n}$.  The 2D mask matrices $\mathbf{M}_s$ and $\mathbf{M}_t$ are governed by the so-called acceleration factor (AF) which indicates the subsampled ratio of the masks, i.e., $AF_s = \frac{n^2}{\Vert\mathbf{M}_s\Vert_0}$ and $AF_t = \frac{n^2}{\Vert\mathbf{M}_t\Vert_0}$. The higher the AF, the fewer k-space scans are employed, leading to faster acquisition. Thus, we consider that $AF_t\ll AF_s$, therefore, the teacher is infeasible in practice since it requires high acquisition time, but it allows high recovery performance, while the student requires reduced acquisition time.
\end{example}

\begin{example}[Compressed Sensing]
Compressed sensing, in particular, the single pixel camera (SPC) acquires coded projections of the scene using coded apertures \cite{duarte2008single}. Here,  the teacher and the student are defined as {$\mathbf{A}_t \in \{-1,1\}^{m_t\times n}$ and $\mathbf{A}_s \in \{-1,1\}^{m_s\times n}$, respectively, where $m_t$ and $m_s$ are the number of snapshots and each row comes from the Hadamard basis. }  Increasing the number of snapshots causes increased acquisition and processing time. Thus, we set that $m_t\gg m_s$ such that the student can be easily implemented in practice. {In this context, the compression ratios are given by $ \gamma_t = \frac{m_t}{n} $ for the teacher and $ \gamma_s = \frac{m_s}{n} $ for the student.}
\end{example}

\begin{example}[Super Resolution]Consider single-image super-resolution, where the goal is to recover a
high-resolution image $\mathbf{x}\in\mathbb{R}^{n}$ from a low-resolution
observation obtained after blur and spatial decimation. Following the standard degradation model, let $\mathbf{D}_s\in\{0,1\}^{m_s\times n}$ and $\mathbf{D}_t\in\{0,1\}^{m_t\times n}$ be downsampling matrices, and $\mathbf{B}_s,\mathbf{B}_t\in\mathbb{R}^{n\times n}$ the blur matrices.   
$\mathbf{A}_s=\mathbf{D}_s\mathbf{B}_s\in\mathbb{R}^{m_s\times n}$ and  
$\mathbf{A}_t=\mathbf{D}_t\mathbf{B}_t\in\mathbb{R}^{m_t\times n}$ with $RF_s=\sqrt{n/m_s}\gg RF_t=\sqrt{n/m_t}$ \cite{tian2011survey}.
\end{example}

Based on these criteria, we want to design the PO such that the DIPA behaves similarly to the virtual teacher algorithm.

\subsection{Distilling the preconditioning operator}
Our approach is data-driven, where a dataset of clean images $\mathcal{X} = \{\mathbf{x}_i\}_{i=1}^{T}$ with $T$ images is employed to generate the student and teacher measurements as $\mathcal{Y}_s = \{\mathbf{y}_{s_i} \vert \mathbf{y}_{s_i} = \mathbf{A}_s\mathbf{x}_i + \mathbf{e}_{s_i}\}_{i=1}^T$ and  $\mathcal{Y}_t = \{\mathbf{y}_{t_i} \vert \mathbf{y}_{t_i} = \mathbf{A}_t\mathbf{x}_i + \mathbf{e}_{t_i}\}_{i=1}^T$. The optimization of the PO is proposed as 
\begin{align}
\label{eq:opt_kd}
   {\mathcal{P}^{\star}}=\ &\underset{{\mathcal{P}}}{\operatorname{arg min}} \ 
   \mathbb{E}_{\mathbf{x}_i,\mathbf{y}_{s_i},\mathbf{y}_{t_i}}
    \mathcal{L}_{KD}\left(\hat{\mathbf{x}}_{s_i}(\mathcal{P}),{\mathbf{A}_s}, {\mathbf{A}_t},{\hat{\mathbf{x}}_{t_i}}, {\mathcal{P}}\right) 
    \nonumber\\
    &\text{ s.t }\hat{\mathbf{x}}_{s_i}({\mathcal{P}})=\texttt{DIPA}\left({\mathcal{P}},{\mathbf{A}_s},{\mathbf{y}_{s_i}}\right)\nonumber\\
    &\hphantom{\text{ s.t }}\hat{\mathbf{x}}_{t_i}=\texttt{TA}\left({\mathbf{A}_t},{\mathbf{y}_{t_i}}\right),
\end{align}

where the notation $\texttt{DIPA}\left({\mathcal{P}},{\mathbf{A}_s},{\mathbf{y}_{s_i}}\right)$ and $\texttt{TA}\left({\mathbf{A}_t},{\mathbf{y}_{t_i}}\right)$ {denote the student and teacher reconstruction algorithms, respectively. Thus, $\hat{\mathbf{x}}_{s_i}(\mathcal{P})$ is the student output after the fixed number of iterations with PO $\mathcal{P}$, and $\hat{\mathbf{x}}_{t_i}$ is the corresponding teacher output.} The crucial aspect here is the cost function $\mathcal{L}_{KD}(\cdot)$ that depends on the teacher and student sensing matrices and reconstructions. The proposed cost function is the following 
\begin{align}
    &\mathcal{L}_{KD}= \mathcal{L}_{G}(\hat{\mathbf{x}}_{s_i}(\mathcal{P}),{\mathbf{A}_s}, {\mathbf{A}_t},{\hat{\mathbf{x}}_{t_i}}, {\mathcal{P}}) + \beta  \mathcal{L}_I(\hat{\mathbf{x}}_{s_i}(\mathcal{P}),{\hat{\mathbf{x}}_{t_i}}),
\end{align}
where the first term, denoted as the gradient loss function $\mathcal{L}_G$, {aligns the direction of the preconditioned student data-fidelity gradient with the teacher data-fidelity gradient} as follows
\begin{equation}
\mathcal{L}_{G}=\left(1- \mathcal{S}_c\left(\mathcal{P}\left(\nabla g_s(\hat{\mathbf{x}}_{s_i})\right), \nabla g_t({\hat{\mathbf{x}}_{t_i}})\right)\right)^2, \label{eq:LG}
\end{equation}
where $g_s(\hat{\mathbf{x}}_{s_i}) = \Vert\mathbf{y}_{s_i}-\mathbf{A}_s\hat{\mathbf{x}}_{s_i}\Vert_2^2$, $g_t(\hat{\mathbf{x}}_{t_i}) = \Vert\mathbf{y}_{t_i}-\mathbf{A}_t\hat{\mathbf{x}}_{t_i}\Vert_2^2$ are the student and teacher data fidelity terms, $\mathcal{S}_c$ is the cosine similarity, i.e., {$\mathcal{S}_c(\mathbf{a},\mathbf{b}) = \frac{{\mathbf{a}^T\mathbf{b}}}{\Vert\mathbf{a}\Vert\Vert\mathbf{b}\Vert}$}. This loss function guides the PO design {so that the preconditioned student data-fidelity gradient has a direction similar to the teacher gradient; in implementation, the cosine denominator is stabilized with a small positive constant when needed.} Finally, the term $\mathcal{L}_I$ is called an imitation loss function since its main objective is that the DIPA obtains the same output as the teacher algorithm. This term is defined as 

\begin{equation}
   \mathcal{L}_I = \Vert\hat{\mathbf{x}}_{s_i}(\mathcal{P})-{\hat{\mathbf{x}}_{t_i}}\Vert_2^2.\label{eq:LI}
\end{equation}

Combining these two loss functions allows the PO design to behave like the teacher model in the DIPA without being highly affected by the physical limitations of the real implementation. The optimization problem is solved using off-the-shelf stochastic gradient descent algorithms such as {Adam \cite{adam} or AdamW \cite{loshchilov2017decoupled}}. Note that unlike traditional PO design, which only aims to improve the algorithm's convergence rate,  our approach aims to improve recovery performance by the guidance of the PnP teacher algorithm with the $\mathcal{L}_I$ loss function and to achieve good convergence rate via the $\mathcal{L}_G$ loss.  

Additionally, the supervised loss, as an alternative to the imitation loss, optimizes the DIPA student's PO using the ground truth label $\mathbf{x}_i$ instead of the teacher's output, defined as 
\begin{equation}
\mathcal{L}_{S}= \Vert\hat{\mathbf{x}}_{s_i}(\mathcal{P})-{{\mathbf{x}_i}}\Vert_2^2.    
\end{equation}
{This loss is used only in the ablation settings where clean training images are available; it is not used as additional information at test time.} 
\subsection{L-DIPA Formulation}
\label{sec:convergence_reg}
L-DIPA formulates the PO as PM where $\mathcal{P} = \mathbf{P}\in\mathbb{R}^{n\times n}$, which allows better interpretability in the recovery algorithm. {Unlike classical symmetric positive-definite preconditioners designed only for solving a fixed linear system, our $\mathbf{P}$ is learned as a gradient transformation inside a reconstruction algorithm. Symmetric parameterizations such as $\mathbf{P}=\mathbf{L}\mathbf{L}^\top$ are possible, but we keep the unconstrained linear form to avoid restricting the teacher-student gradient matching; the regularizer below penalizes mismatch with the teacher update.} Particularly, we can exploit the convergence criteria of the teacher to regularize the design of $\mathbf{P}$. Here we focus on the PnP-FISTA scheme with a linear PO $\mathbf{P}$; This analysis can be adapted to the RED-FISTA scheme as well. 
\begin{assumption}[Bounded denoiser]
The denoiser $\mathrm{D}_\sigma$ is bounded such that for $\mathbf{x},\mathbf{z}\in\mathbb{R}^n$
\begin{equation}
    \Vert\mathrm{D}_\sigma(\mathbf{x})-\mathrm{D}_\sigma(\mathbf{z})\Vert_2 \leq (1+\delta)\Vert\mathbf{x-z}\Vert_2,
\end{equation}
where $\delta>0$ is a numerical constant.
\end{assumption}

Usually, the convergence of the PnP algorithm is derived via fixed point analysis \cite{chan2016plug}. {Here, we derive a convergence-motivated discrepancy term between one student update and the corresponding teacher update; the term is used only as a training regularizer, not as a formal proof that the student has the same fixed point as the teacher.} First, consider the iterations given by

\begin{equation}
    \mathbf{x}_s^{k} =  \mathrm{T}_s(\mathbf{x}_s^{k-1}) = \mathrm{D}_\sigma(\mathbf{x}_s^{k-1} - \alpha \mathbf{P}(\mathbf{A}_s^\top (\mathbf{A}_s\mathbf{x}_s^{k-1} - \mathbf{y}_s))), \nonumber
\end{equation}
and
\begin{equation}
    \mathbf{x}_t^{k} =  \mathrm{T}_t(\mathbf{x}_t^{k-1}) = \mathrm{D}_\sigma(\mathbf{x}_t^{k-1} - \alpha (\mathbf{A}_t^\top (\mathbf{A}_t\mathbf{x}_t^{k-1} - \mathbf{y}_t))). \nonumber
\end{equation}
{Under the noiseless training model used in this bound, i.e., $\mathbf{y}_s=\mathbf{A}_s\mathbf{x}$ and $\mathbf{y}_t=\mathbf{A}_t\mathbf{x}$, we compare the two one-step updates as}
\begin{equation}
\begin{aligned}
\left\|\mathbf{x}_{s}^{k} - \mathbf{x}_t^{k}\right\|  &=
\left\|\mathrm{T}_{s}\left({\mathbf{x}_s^{k-1}}\right) - \mathrm{T}_t\left({\mathbf{x}_t^{k-1}}\right)\right\|\\&= \Big\|\mathrm{D}_\sigma\Big({\mathbf{x}_s^{k-1}} - \alpha \mathbf{P} \mathbf{A}_s^\top \big(\mathbf{A}_s {\mathbf{x}_s^{k-1}} - \overbrace{\mathbf{A}_s \mathbf{x}}^{\mathbf{y}_s}\big)\Big)  - \mathrm{D}_\sigma\Big({\mathbf{x}_t^{k-1}} - \alpha \mathbf{A}_t^\top \big(\mathbf{A}_t {\mathbf{x}_t^{k-1}} - \underbrace{\mathbf{A}_t \mathbf{x}}_{\mathbf{y}_t}\big)\Big)\Big\| \\
&\leq (1 + \delta) \left\|\left(\mathbf{I} - {\alpha}\mathbf{P} \mathbf{A}_s^\top \mathbf{A}_s\right) {\mathbf{x}_s^{k-1}} 
- \left(\mathbf{I} - {\alpha}\mathbf{A}_t^\top \mathbf{A}_t\right) {\mathbf{x}_t^{k-1}}\right\|  + {R_C(\mathbf{P},\mathbf{A}_s,\mathbf{A}_t,\mathbf{x})},
\end{aligned}
\end{equation}
where $\mathbf{x}$ is the ground truth image, the third line employs {Assumption 1}, and the fourth line follows the triangle inequality. {Here, $R_C(\mathbf{P},\mathbf{A}_s,\mathbf{A}_t,\mathbf{x})=\alpha\| (\mathbf{P}\mathbf{A}_s^\top\mathbf{A}_s-\mathbf{A}_t^\top\mathbf{A}_t)\mathbf{x}\|$ is the proposed convergence regularization term; it encourages the linearized student data-fidelity update to match the teacher update on the training image manifold.} We choose this function as regularization since it only depends on the ground truth image available in training and the student and teacher sensing matrices. The optimization problem based on this {formulation} becomes 
\begin{align}
\label{eq:opt_reg}
   {\mathbf{P}^{\star}}=\ &\underset{{\mathbf{P}}}{\operatorname{arg min}} \ 
   \mathbb{E}_{\mathbf{x}_i,\mathbf{y}_{s_i},\mathbf{y}_{t_i}}
    \mathcal{L}_{KD}\left(\hat{\mathbf{x}}_{s_i}(\mathbf{P}),{\mathbf{A}_s}, {\mathbf{A}_t},{\hat{\mathbf{x}}_{t_i}}, {\mathbf{P}}\right) + {\tau R_C(\mathbf{P},\mathbf{A}_s,\mathbf{A}_t,\mathbf{x})}
    \nonumber\\
    &\text{ s.t }\hat{\mathbf{x}}_{s_i}({\mathbf{P}})=\texttt{DIPA}\left({\mathbf{P}},{\mathbf{A}_s},{\mathbf{y}_{s_i}}\right)\nonumber\\
    &\hphantom{\text{ s.t }}\hat{\mathbf{x}}_{t_i}=\texttt{TA}\left({\mathbf{A}_t},{\mathbf{y}_{t_i}}\right),
\end{align}

\subsection{N-DIPA Formulation}
{For higher-dimensional inverse problems, replacing the dense matrix $\mathbf{P}$ with a convolutional neural operator $\mathcal{P}=\mathcal{P}_\theta$ makes the parameter count independent of the image dimension. Although both sensing models are linear, the desired map from a student gradient to a teacher-like update is generally state-, prior-, and iteration-dependent once the denoiser and finite unrolled reconstruction are included. A nonlinear PO can therefore model a data-dependent gradient transformation that a single global matrix may not capture.} To improve this preconditioning, we incorporate an iteration-aware scheme for the PO that adapts depending on the iteration without significantly increasing the number of parameters. We use a positional encoding $\phi(k)$ that lifts the iteration index into a {higher-dimensional} sinusoidal space, and these frequencies are used in the operator $\mathcal{P}_\theta$. {Algorithm~\ref{alg:npo_fista} shows PnP-FISTA with N-DIPA, and Algorithm~\ref{alg:kd_npo} shows the corresponding PO training procedure.} 

\begin{algorithm}[H]
\footnotesize 
\caption{Non-Linear Preconditioned FISTA}
\label{alg:npo_fista}
\begin{algorithmic}[1]
\Require $\mathcal{P},K,\mathbf{A},\mathbf{y},\alpha,\lambda$
  \State $\mathbf{x}^0=\mathbf{z}^1=\mathbf0,\;t =1$
  \For{$k=1,\dots,K$}
    \State $\mathbf{x}^k\gets\mathbf{z}^k-\alpha\mathcal{P}(\nabla g(\mathbf{z}^k),\phi(k))$
    \State $\mathbf{x}^k\gets\operatorname{prox}_{\gamma\lambda h}(\mathbf{x}^k)$
    \State $t^\prime = t$
    \State $t = \frac{1+\sqrt{1+4t^\prime)^2}}{2}$
    \State $\mathbf{z}^{k+1}\gets\mathbf{x}^k+\frac{t^\prime-1}{t}(\mathbf{x}^k-\mathbf{x}^{k-1})$
  \EndFor
  \State\Return $\mathbf{x}^K$
\end{algorithmic}
\end{algorithm}
    
\begin{algorithm}[H]
\footnotesize 
\caption{KD Training of Nonlinear Preconditioner}
\label{alg:kd_npo}
\begin{algorithmic}[1]
\Require $N, \mathcal{X}, \mathcal{Y}_s,\mathcal{Y}_t,\mathbf{A}_s, \mathbf{A}_t,\alpha_s,\alpha_t{,\beta_I,\beta_S}$
\State Initialize randomly $\theta$ 
\For{$epoch = 1,\dots,N$}
  \For{ $i = 1,\dots, B$}
    \State $\hat{\mathbf{x}}_{{\mathcal{P}_\theta s}_i} = \texttt{Algorithm 1} (\mathcal{P}_\theta, K, \mathbf{A}_s,\mathbf{y}_{s_i}, \alpha_s,\lambda)$
    \State $\hat{\mathbf{x}}_{t_i} = \texttt{Algorithm 1} (\mathcal{I}, K, \mathbf{A}_t,\mathbf{y}_{t_i}, \alpha_t,\lambda)$
    \State $\mathcal{L}_G=\left|1-\tfrac{\mathcal{P}_\theta(\nabla g_s(\hat{\mathbf{x}}_{{\mathcal{P}_\theta s}_i} ))^\top{g}_t(\hat{\mathbf{x}}_{t_i} )}{\|\mathcal{P}_\theta(\nabla g_s(\hat{\mathbf{x}}_{{\mathcal{P}_\theta s}_i} ))\|\|{g}_t(\hat{\mathbf{x}}_{t_i} )\|}\right|^2$
    \State $\mathcal{L}_I=\|\hat{\mathbf{x}}_{{\mathcal{P}_\theta s}_i} -\hat{\mathbf{x}}_{t_i}\|_2^2$
    \State $\mathcal{L}_S=\|\hat{\mathbf{x}}_{{\mathcal{P}_\theta s}_i} -\mathbf{x}_i\|_2^2$
    \State $\theta\gets\theta-\eta\,\nabla_\theta\bigl(\mathcal{L}_G{+\beta_I\,\mathcal{L}_I+\beta_S\,\mathcal{L}_S}\bigr)$
  \EndFor
\EndFor
\end{algorithmic}

\end{algorithm}

\section{Experiments}

We validate DIPA on three imaging inverse problems: single-coil MRI, super-resolution (SR), and compressed sensing (CS) with a single-pixel camera (SPC). Unless otherwise stated, all reconstructions use the RED-FISTA or PnP-FISTA implementations from \cite{Tachella_DeepInverse_A_deep_2023}. The convergence regularization parameter is fixed to $\tau=1\times 10^{-3}$ for both RED and PnP, and the unpreconditioned baseline corresponds to initializing and keeping the PO as the identity operator. We use $20$ FISTA iterations with teacher and student stepsizes $\alpha_t=0.7$ and $\alpha_s=0.4$, respectively, and optimize each PO for $50$ epochs. Experiments were run on an Intel(R) Xeon(R) W-3223 CPU @ 3.50GHz, 48 GB RAM, and an NVIDIA GeForce RTX 3090 with 24 GB VRAM.

\textbf{MRI:} We use the FastMRI single-coil knee dataset \cite{FastMRI_dataset}, preprocessed by \cite{Tachella_DeepInverse_A_deep_2023}. The split contains $900$ training and $73$ testing images. The original $320\times320$ images are resized to $50\times50$ ($n=2500$) {only to keep the dense linear PO tractable; this limitation motivates N-DIPA and the additional $128\times128$ CS ablations below.} We use 1D Gaussian undersampling masks, AdamW \cite{loshchilov2017decoupled} with learning rate $1\times10^{-5}$ and weight decay $0.01$, and batch sizes of $8$ for RED and $32$ for PnP. 

\textbf{Compressed sensing:} We use MNIST \cite{deng2012mnist} with $50,000$ training and $10,000$ testing images resized to $32\times32$. The SPC sensing matrix is a 2D subsampled Hadamard transform. We optimize with Adam \cite{adam}, learning rate $1\times10^{-5}$, and batch sizes of $50$ for RED and $225$ for PnP.

\textbf{Super-resolution:} We use CelebA \cite{liu2015faceattributes} images resized to $110\times110$. The PO is optimized with Adam \cite{adam} and learning rate $1\times10^{-5}$. Additional ablations for SR, higher-resolution CS, E2E comparisons, and robustness checks are integrated in Section \ref{app:abblation_loss}--Section \ref{app:Cross_validation}.

\subsection{State-of-the-art comparison}

Table \ref{tab:SOTA_SR_MRI_CS} compares DIPA with classical and learning-based preconditioning methods under the same reconstruction framework. The non-learned Hessian and polynomial preconditioners provide modest gains over the identity baseline, while existing learned preconditioners improve some tasks but are not consistently strong across MRI, SR, and CS. L-DIPA gives the best linear-preconditioner result in all three settings, and N-DIPA gives the strongest overall PSNR: $29.39$ dB for MRI, $25.88$ dB for SR, and $34.55$ dB for CS.

\begin{figure}[pos=!t]

    \centering
    \includegraphics[width=\linewidth]{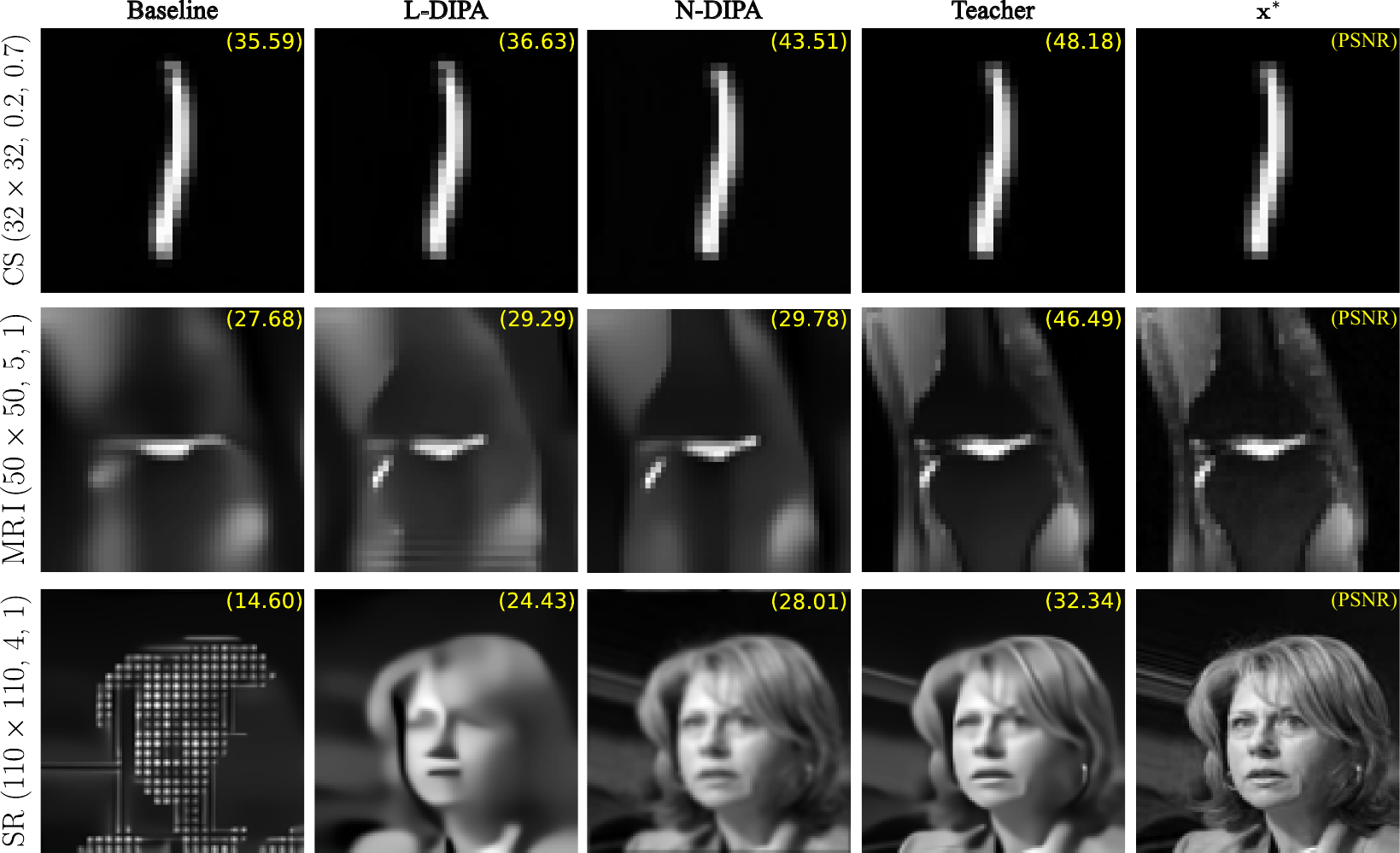}
    \caption{Visual results and PSNR for PnP-FISTA with different preconditioning methods in SPC, MRI, and SR. 
    } 
    \label{fig:D2GP_VisualResults}
\end{figure}

The qualitative results in Fig. \ref{fig:D2GP_VisualResults} follow the same trend as the quantitative comparison: DIPA reduces artifacts relative to the baseline and closes much of the gap to the teacher reconstruction. Fig. \ref{fig:SR_convergence} further shows that both L-DIPA and N-DIPA improve the convergence behavior over the competing preconditioners on SR.

\subsection{L-DIPA results}

Fig. \ref{fig:POs} visualizes the learned linear preconditioning matrices. Starting from the identity, L-DIPA learns task-dependent structures: CS produces block-like and diagonal patterns, MRI emphasizes diagonal regions, and SR produces vertical structures along the diagonal. These differences indicate that the learned PM adapts to the sensing model rather than acting only as a generic convergence accelerator.

\begin{figure}[pos=!t]
    \centering
    \includegraphics[width=0.9\linewidth]{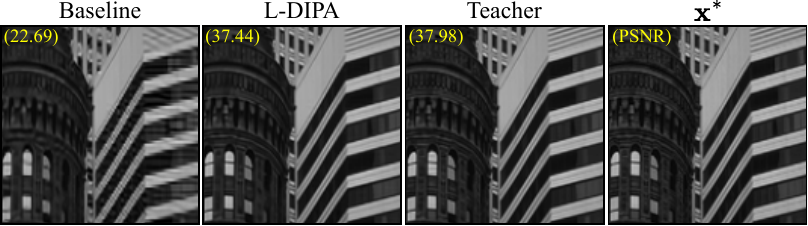}
    \caption{Visual results and PSNR for Compressive Sensing $(128 \times 128)$ with RED-FISTA, $\gamma_t=0.7$, $\gamma_s=0.2$, $\mathcal{R}_{C}(\text{\xmark})$, $\mathcal{L}_{S}(\text{\cmark})$ with the BSDS500 dataset \cite{arbelaez2010contour}.}
    \label{fig:cs_128_RED}
\end{figure}

The additional RED-FISTA CS result in Fig. \ref{fig:cs_128_RED} complements the main PnP-FISTA comparison. On $128\times128$ BSDS500 images, L-DIPA substantially improves over the baseline and approaches the teacher reconstruction, with gains of up to approximately $15$ dB in PSNR for the shown setting.

\subsection{N-DIPA results}

For N-DIPA, we use a ConvNeXt network \cite{liu2022convnet} with $5$ blocks and $128$ features per block, a residual connection in the last layer, ReduceLROnPlateau scheduling, and the positional encoding from IndiUNet \cite{delbracio2023inversion} so that the network can condition the preconditioned gradient on the FISTA iteration. The network is optimized for $50$ epochs with Adam and learning rate $1\times10^{-5}$. The architecture selection study in Section \ref{app:NN_selection} supports this choice.

\begin{figure}[pos=!t]
    \centering
    \includegraphics[width=\linewidth]{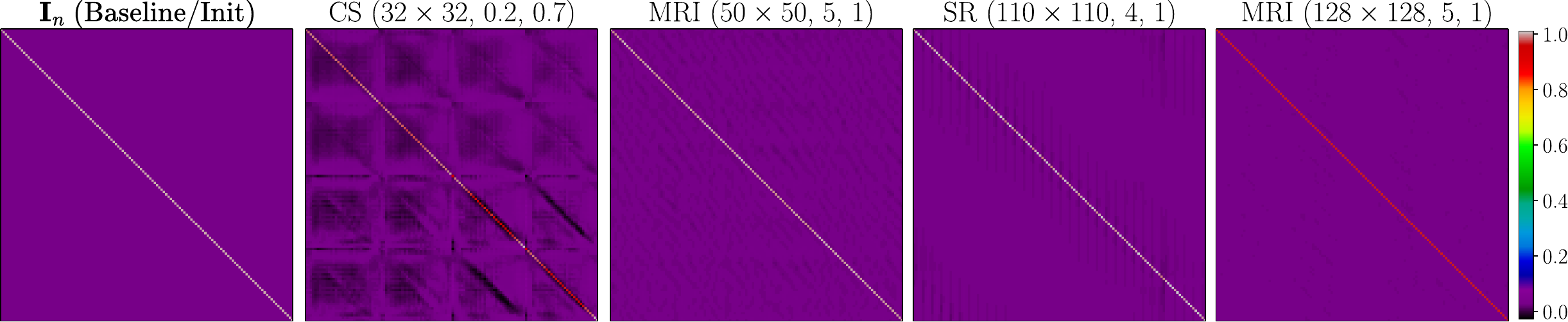}
    \caption{Natural logarithmic representation of the linear PO, $\log(PO+1)$, for different tasks and resolutions.}
    \label{fig:POs}
\end{figure}

\begin{figure}[pos=!t]
  \centering
    \centering
    \includegraphics[width=0.6\linewidth]{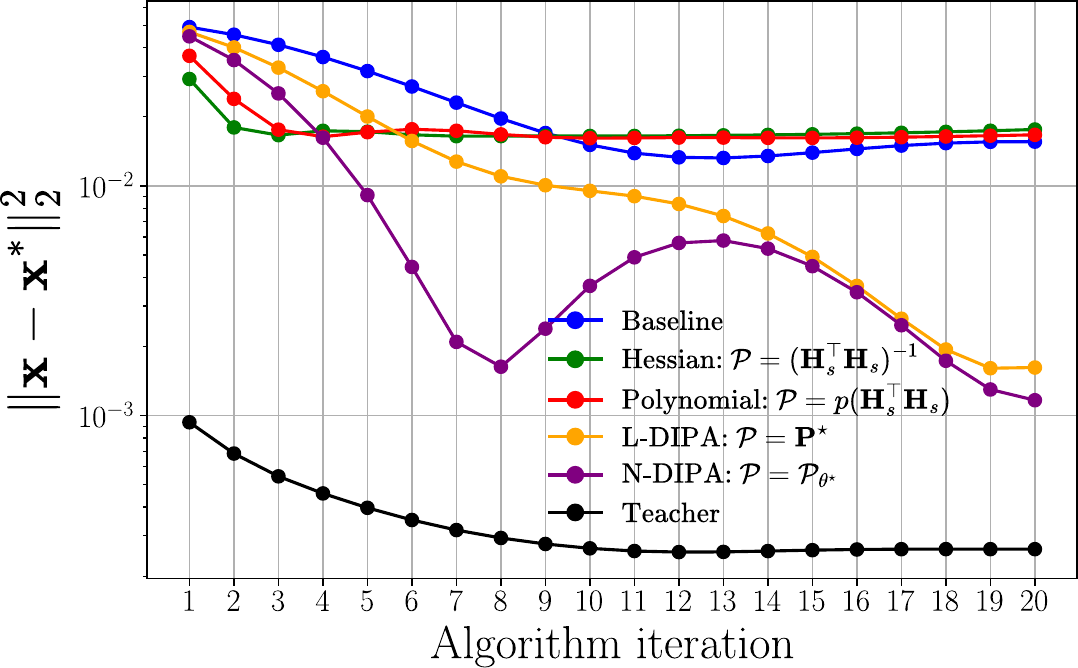}
    \caption{Signal convergence with state-of-the-art preconditioning methods for SR task.}
    \label{fig:SR_convergence}
\end{figure}

N-DIPA improves over L-DIPA in all tasks in Table \ref{tab:SOTA_SR_MRI_CS}, with the largest advantage in CS, where the nonlinear PO reaches $34.55$ dB compared with $28.10$ dB for L-DIPA. {The N-DIPA curve in Fig.~\ref{fig:SR_convergence} is not strictly monotone, which is expected for a learned nonlinear PO composed with a denoiser and evaluated with a fixed unrolled solver; therefore, we use the fixed-iteration reconstruction quality, not monotonicity of every intermediate iterate, as the primary metric.} To connect this nonlinear behavior with the interpretable linear case, Section \ref{app:linearization} provides a finite-difference linearization of the trained N-DIPA operators; the resulting matrices show structures similar to the L-DIPA PMs in Fig. \ref{fig:POs}.
\begin{table*}[!t]
\centering
\resizebox{\linewidth}{!}{
\begin{tabular}{c|c|c|c|c|c}\hline\hline
Learning&Method & Preconditioning model & MRI ($AF_s=5$) &  SR ($RF_s=4$) & CS ($\gamma_s = 0.2$)   \\\hline
\xmark&Baseline & $ \mathbf{I}_n $ & $25.77$ & $11.14$ & 22.36\\
\xmark&Hessian~\cite{dassios2015preconditioner,fessler1999conjugate} &   $ (\mathbf{A}^\top\mathbf{A})^{-1}$& {28.16} & 11.75 & 22.94 \\
\xmark &Polynomial~\cite{iyer2024polynomial,tan2024provably,johnson1983polynomial} &  $p(\mathbf{A}^\top\mathbf{A})$  & 28.07 & 11.81 & 23.01 \\
\cmark&Scalar step~\cite{ehrhardt2024learning} & $ p_k \mathbf{I_n} $ & $27.88$  & 11.62 & 23.36 \\
\cmark&Convolutional~\cite{ehrhardt2024learning} & $ \mathbf{p_k \ast x} $ & {$28.63$}  & 17.55 & 21.30\\
\cmark&Pointwise~\cite{ehrhardt2024learning} & $ \mathbf{p_k \odot x} $ & $ 27.59$  &  11.19 & 21.28\\
\cmark&Full-linear~\cite{ehrhardt2024learning}& $ \mathbf{P_k} $ (see \cite{ehrhardt2024learning}) & {$28.15$}  & {22.02}  & {26.19}  \\
\cmark&L-DIPA (Ours) & $\mathbf{P}^{\star} \eqref{eq:opt_reg} $ & $\underline{29.21}$ & $\underline{25.00}$ & $\underline{28.10}$ \\
\cmark&N-DIPA (Ours) & $\mathcal{P}_{\theta^{\star}} \eqref{eq:opt_reg} $ & $\underline{\textbf{29.39}}$ & $\underline{\textbf{25.88}}$ & $\underline{\textbf{34.55}}$ \\
\hline \hline
\end{tabular}
}
\vspace{-0.3cm}
\caption{Preconditioning comparison with PnP-FISTA for MRI, SR, and CS.}
\label{tab:SOTA_SR_MRI_CS}
\end{table*}


\subsection{Convergence regularization and supervised loss function ablation studies} \label{app:abblation_loss}

Tables \ref{tab:GP-PnP} and \ref{tab:GP-RED} analyze the influence of the supervised loss $\mathcal{L}_{S}$, the convergence regularizer $R_C$, and the PnP/RED prior choice for MRI and CS. In both modalities, learning the PO improves over the identity baseline; the gains reach almost 3 dB in MRI and 10 dB in CS. L-DIPA PnP gives the strongest improvement in these ablations.

\begin{figure}[pos=!t]
    \centering
     \includegraphics[width=0.6\columnwidth]{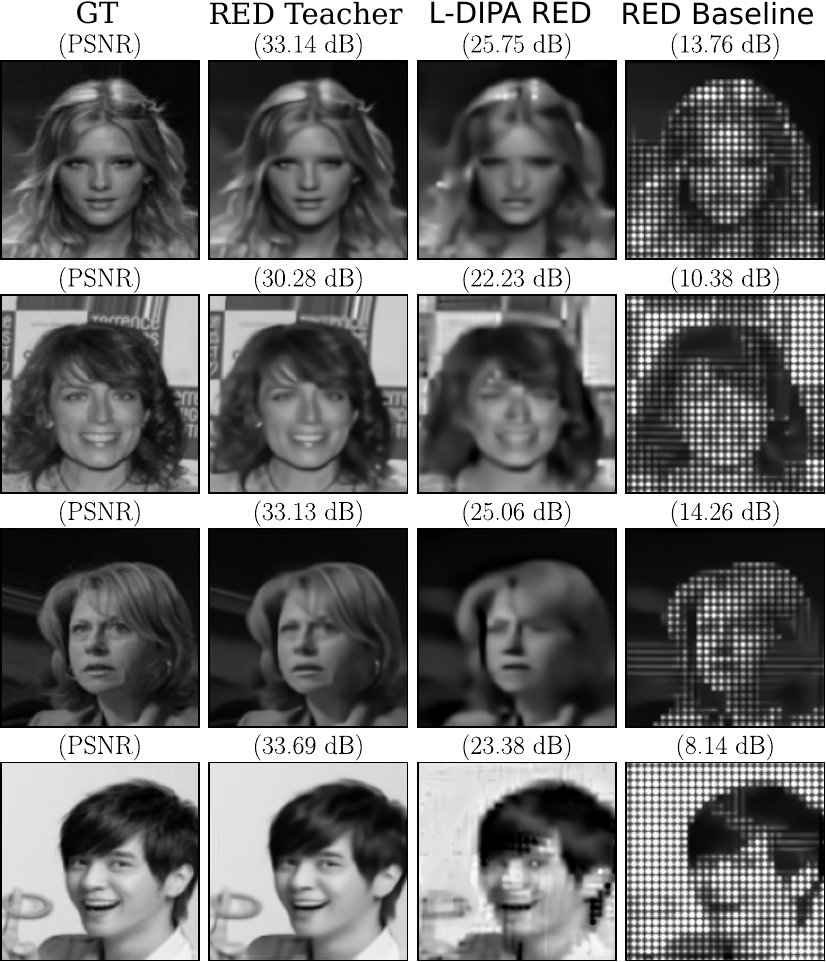}
     \vspace{-0.1cm}
    \caption{Visual results and PSNR for Super Resolution $(110 \times 110)$ with RED-FISTA, $RF_t=1$, $RF_s=4$, $\mathcal{R}_{C}(\text{\cmark})$, $\mathcal{L}_{S}(\text{\xmark})$ with the CelebA dataset \cite{liu2015faceattributes}.}
    \label{fig:sr_110_RED}
\end{figure}

Complementing the MRI and CS ablations, we evaluate SR on CelebA \cite{liu2015faceattributes} with images resized to $110 \times 110$. We fix $RF_s=4$, vary the teacher configuration with $RF_t=\{1,2,3\}$, and optimize for 50 Adam iterations \cite{adam} with learning rate $1\times 10^{-5}$. Table \ref{tab:GP-SuperResolution} shows PSNR gains of up to 14 dB from teacher guidance. In some settings, such as $\mathcal{R}_C(\textbf{\cmark})$ and $\mathcal{L}_S(\textbf{\cmark})$ for $RF_t=3$, the student outperforms the teacher because the supervised loss uses ground truth during training. Fig. \ref{fig:sr_110_RED} shows the corresponding qualitative behavior: L-DIPA RED is much closer to the teacher than to the baseline despite using a four-times lower-resolution measurement. We also include a higher-resolution CS ablation on BSDS500 \cite{arbelaez2010contour}, resized to 128$\times$128 in grayscale. The training settings match the main CS comparison in Table \ref{tab:SOTA_SR_MRI_CS}. Table \ref{tab:CS-128-BSDS500} shows improvements of up to 4 dB in PSNR. {Table~\ref{tab:CS-128-MNIST} reports the corresponding 128$\times$128 MNIST ablation, showing that the teacher-guided PO remains beneficial after upsampling the digits.}

\begin{table}[!h] 
\centering
\begin{tblr}{
  cells = {c},
  cell{1}{3} = {c=3}{},
  cell{1}{6} = {c=3}{},
  cell{3}{1} = {r=3}{},
  cell{3}{2} = {r=3}{},
  cell{6}{1} = {r=3}{},
  cell{6}{2} = {r=3}{},
  vline{3,6,9} = {1}{},
  vline{-} = {2-8}{},
  vline{4-9} = {4-5,7-8}{},
  vline{4-5,7-8} = {9-10}{},
  hline{1} = {3-8}{},
  hline{2-3,6,9} = {-}{},
  hline{11} = {4,7}{},
}
&& \textbf{MRI} &&& \textbf{CS} && \\
$\mathcal{R}_{C}$ & $\mathcal{L}_{S}$ & $AF_t$ & \textbf{L-DIPA PnP} & \textbf{Teacher} & $\gamma_t$ & \textbf{L-DIPA PnP} & \textbf{Teacher} \\

\xmark & \cmark & 3 & 28.54 & 29.60 & 0.5 & 32.73 & 34.65 \\
&& 2 & \uline{28.56} & 40.45 & 0.7 & \uline{32.74} & 39.89 \\
&& 1 & 28.54 & 48.66 & 0.9 & 32.73 & 50.06 \\

\cmark & \xmark & 3 & 26.88 & 29.60 & 0.5 & 31.88 & 34.65 \\
&& 2 & 28.53 & 40.45 & 0.7 & 32.61 & 39.89 \\
&& 1 & 28.53 & 48.66 & 0.9 & 32.71 & 50.06 \\

&&& \textbf{Base.} &&& \textbf{Base.} & \\
&&& 25.77 &&& 22.88 &
\end{tblr}
\caption{Ablation study in PSNR for MRI and CS with different $AF_t$ values and $\gamma_t$ values for the {PnP} teacher, with $AF_s=5$ and $\gamma_s=0.2$ for the student (L-DIPA PnP), respectively. {The baseline PnP is the case when the PM is $\mathbf{P}=\mathbf{I}$.} }
\label{tab:GP-PnP}
\end{table}

\begin{table}[!h] 
\centering
\begin{tblr}{
  cells = {c},
  cell{1}{3} = {c=3}{},
  cell{1}{6} = {c=3}{},
  cell{3}{1} = {r=3}{},
  cell{3}{2} = {r=3}{},
  cell{6}{1} = {r=3}{},
  cell{6}{2} = {r=3}{},
  vline{3,6,9} = {1}{},
  vline{-} = {2-8}{},
  vline{4-9} = {4-5,7-8}{},
  vline{4-5,7-8} = {9-10}{},
  hline{1} = {3-8}{},
  hline{2-3,6,9} = {-}{},
  hline{11} = {4,7}{},
}
&& \textbf{MRI} &&& \textbf{CS} && \\
$\mathcal{R}_{C}$ & $\mathcal{L}_{S}$ & $AF_t$ & \textbf{L-DIPA RED} & \textbf{Teacher} & $\gamma_t$ & \textbf{L-DIPA RED} & \textbf{Teacher} \\

\xmark & \cmark & 3 & 30.5 & 28.75 & 0.5 & 31.51 & 32.99 \\
&& 2 & 27.59 & 43.76 & 0.7 & \uline{31.51} & 37.51 \\
&& 1 & 30.39 & 53.00 & 0.9 & 31.50 & 50.64 \\

\cmark & \xmark & 3 & 27.68 & 28.75 & 0.5 & 30.31 & 32.99 \\
&& 2 & 27.54 & 43.76 & 0.7 &  31.10 & 37.51 \\
&& 1 & 27.58 & 53.00 & 0.9 &  31.21 & 50.64 \\

&&& \textbf{Base.} &&& \textbf{Base.} & \\
&&& 27.31 &&& 22.93 &
\end{tblr}
\caption{Ablation study in PSNR for MRI and CS with different $AF_t$ values and $\gamma_t$ values for the {RED} teacher, with $AF_s=5$ and $\gamma_s=0.2$ for the student (L-DIPA RED), respectively. {The baseline RED is the case when the PO $\mathcal{P}=\mathbf{I}$.}}
\label{tab:GP-RED}
\end{table}

\begin{table}
\centering

\begin{tblr}{
  cells = {c},
  cell{6}{1} = {c=2}{},
  cell{7}{1} = {c=2}{},
  vline{-} = {1-5}{},
  vline{1,3,5} = {6-7}{},
  hline{1-2,6-8} = {-}{},
}
$\mathcal{R}_{C}$ & $\mathcal{L}_{S}$ & L-DIPA PnP         & L-DIPA RED         \\
\cmark    & \cmark & \textbf{32.65} & 33.45          \\
\xmark    & \cmark & \uline{32.64}  & \textbf{34.99} \\
\cmark    & \xmark & 32.29          & 32.82          \\
\xmark    & \xmark & 32.28          & \uline{33.94}  \\
Teacher  &       & 38.54          & 38.73          \\
Baseline &       & 29.33          & 29.05          
\end{tblr}

\vspace{-0.2cm}
\caption{Ablation study in terms of PSNR for Compressive Sensing $(128 \times 128)$ with $\gamma_t=0.7$ for the {PnP and RED} teachers and $\gamma_s=0.2$ for the students (L-DIPA PnP and L-DIPA RED) with the BSDS500 dataset \cite{arbelaez2010contour}.}
\label{tab:CS-128-BSDS500}
\end{table}

\begin{table}[!t] 
\centering

\begin{tblr}{
  cells = {c},
  cell{1}{3} = {c=3}{},
  cell{3}{1} = {r=3}{},
  cell{3}{2} = {r=3}{},
  cell{6}{1} = {r=3}{},
  cell{6}{2} = {r=3}{},
  cell{9}{1} = {r=3}{},
  cell{9}{2} = {r=3}{},
  cell{12}{1} = {r=3}{},
  cell{12}{2} = {r=3}{},
  vline{3,6} = {1}{},
  vline{-} = {2-14}{},
  vline{4-5} = {14-16}{},
  hline{17} = {4}{},
  hline{1} = {3-6}{},
  hline{2-3,6,9,12,15} = {-}{},
}
&& \textbf{Compressive Sensing} && \\
$\mathcal{R}_{C}$ & $\mathcal{L}_{S}$ & $\gamma_t$ & \textbf{L-DIPA PnP} & \textbf{Teacher-PnP} \\

\cmark & \cmark & 0.5 & 36.28 & 34.65 \\
               && 0.7 & 36.29 & 51.27 \\
               && 0.9 & \textbf{36.32} & 54.23 \\

\xmark & \cmark & 0.5 & 36.30 & 34.65 \\
               && 0.7 & 36.29 & 51.27 \\
               && 0.9 & 36.30 & 54.23 \\

\cmark & \xmark & 0.5 & 33.15 & 34.65 \\
               && 0.7 & 36.30 & 51.27 \\
               && 0.9 & 36.30 & 54.23 \\

\xmark & \xmark & 0.5 & 33.18 & 34.65 \\
               && 0.7 & 36.26 & 51.27 \\
               && 0.9 & 36.30 & 54.23 \\

&&& \textbf{Base.} & \\
&&& 31.51 & 
\end{tblr}

\vspace{-0.3cm}
\caption{Ablation study in terms of PSNR for Compressive Sensing with different $\gamma_t$ values for the PnP Teacher, with $\gamma_s=0.2$ for the L-DIPA PnP with an upsampled to $(128 \times 128)$ MNIST dataset \cite{deng2012mnist}.}
\label{tab:CS-128-MNIST}
\end{table}

\begin{table}[!t] 
\centering
\begin{tblr}{
  cells = {c},
  cell{1}{3} = {c=6}{},
  cell{3}{1} = {r=3}{},
  cell{3}{2} = {r=3}{},
  cell{6}{1} = {r=3}{},
  cell{6}{2} = {r=3}{},
  cell{9}{1} = {r=3}{},
  cell{9}{2} = {r=3}{},
  cell{12}{1} = {r=3}{},
  cell{12}{2} = {r=3}{},
  vline{3,9} = {1}{},
  vline{-} = {2-14}{},
  vline{4-9} = {4-5,7-8}{},
  vline{4-5,7-8} = {15-16}{},
  hline{1} = {3-8}{},
  hline{2-3,6,9,12,15} = {-}{},
  hline{17} = {4,7}{},
}
&& \textbf{Super-Resolution} && \\
$\mathcal{R}_{C}$ & $\mathcal{L}_{S}$ & $RF_t$ & \textbf{L-DIPA PnP} & \textbf{Teacher-PnP} & $RF_t$ & \textbf{L-DIPA RED} & \textbf{Teacher-RED} \\

\cmark & \cmark & 3 & 24.90 & 17.89 & 3 & 24.76 & 17.93 \\
               && 2 & 24.95 & 30.08 & 2 & 24.75 & 30.45 \\
               && 1 & \textbf{25.00} & 33.97 & 1 & 24.81 & 34.49 \\

\xmark & \cmark & 3 & 24.98 & 17.89 & 3 & 24.77 & 17.93 \\
               && 2 & 24.95 & 30.08 & 2 & 24.77 & 30.72 \\
               && 1 & \uline{24.99} & 33.97 & 1 & \uline{24.82} & 34.49 \\

\cmark & \xmark & 3 & 17.56 & 17.89 & 3 & 17.90 & 18.09 \\
               && 2 & 24.51 & 30.08 & 2 & 24.55 & 30.58 \\
               && 1 & 24.92 & 33.97 & 1 & \textbf{25.12} & 34.48 \\

\xmark & \xmark & 3 & 17.58 & 17.90 & 3 & 17.73 & 17.93 \\
               && 2 & 24.59 & 30.08 & 2 & 24.43 & 30.72 \\
               && 1 & 24.89 & 33.97 & 1 & 24.72 & 34.49 \\

&&& \textbf{Base.} &&& \textbf{Base.} & \\
&&& 11.02 &&& 10.83 &
\end{tblr}
\vspace{-0.4cm}
\caption{Ablation study in terms of PSNR for Super-Resolution $(110 \times 110)$ with different $RF_t$ values for the {PnP and RED} teachers, with $RF_s=4$ for the students (L-DIPA PnP and L-DIPA RED) with the CelebA dataset \cite{liu2015faceattributes}.}
\label{tab:GP-SuperResolution}
\end{table}

\subsection{Neural Network ablation studies} \label{app:NN_selection}

\begin{figure}[pos=!t]
  \centering
    \includegraphics[width=\linewidth]{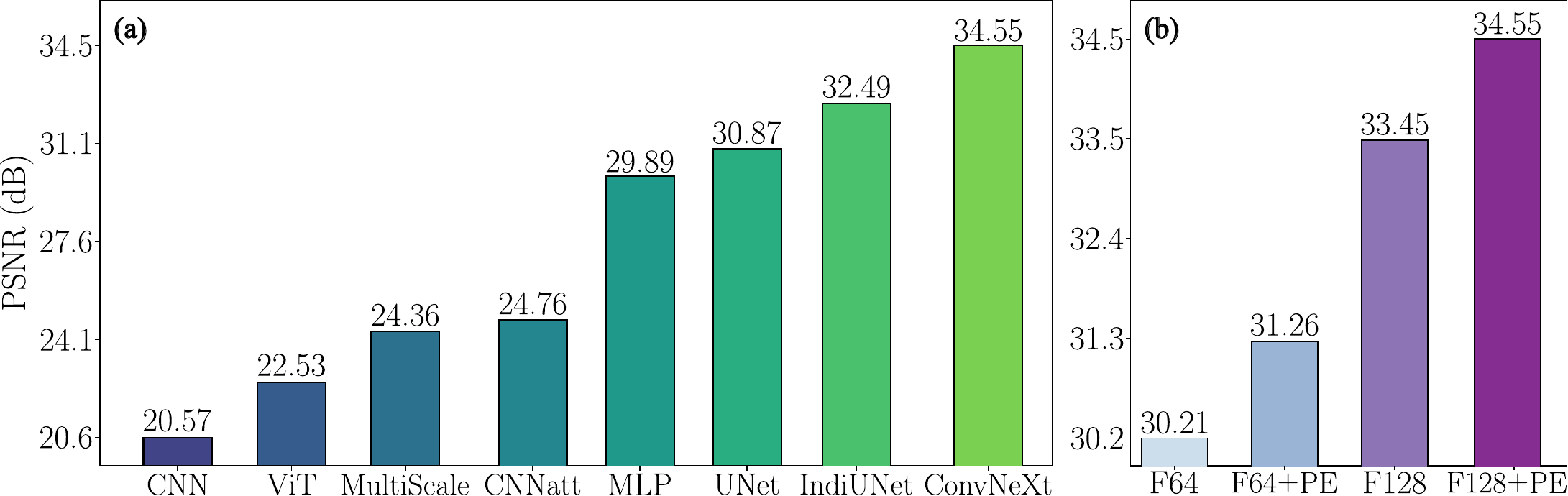}
    \vspace{-0.4cm}
    \caption{\hspace{2mm} (a) Ablation results in terms of PSNR of different NNs for SPC. (b) Number of features and positional encoding usage in ConvNeXt.}
    \label{fig:networks_ablation}
\end{figure}

{This experiment aims to identify the best neural network architecture for the N-DIPA.} {We evaluated the following neural networks (NNs) architectures: } Multilayer perceptron \cite{rosenblatt1958perceptron}, convolutional NN (CNN) \cite{lecun1998gradient}, CNN with attention module \cite{woo2018cbam}, UNet \cite{ronneberger2015u}, MultiScale CNN \cite{yuan2018multiscale}, IndiUNet \cite{delbracio2023inversion}, ViT \cite{dosovitskiy2020image}, and ConvNext \cite{liu2022convnet}. For the ablation studies on Fig. \ref{fig:networks_ablation}(a), modifications were made to reduce the number of parameters of each of the NNs, to have fewer parameters than $\mathbf{P}$, and using the configuration that obtained the best possible performance. Because of this, some networks, such as ViT, despite their widely known potential, are not able to generalize the gradient mapping from student to teacher. Regardless of the configuration, the MLP has more parameters than $\mathbf{P}$, but it does not perform adequately. {ConvNeXt gives the best trade-off in this ablation, likely because its convolutional structure preserves spatial locality while keeping the parameter count below the dense linear PO.}

\subsection{E2E methods {comparison}} \label{app:E2E_ablation}

This comparison is included to assess whether the proposed preconditioner retains the model-adaptation advantages of PnP methods while improving reconstruction quality.
Although E2E methods achieve good recovery performance, they require large datasets and offer limited generalization to data or sensing model variations. In contrast, PnP methods decouple the image formation model from learned denoisers, enabling accurate reconstruction under changing conditions. They have also proven ideal for applications with scarce data and varying acquisition settings \cite{kamilov2023plug}. {DIPA is not intended to replace all end-to-end (E2E) solvers. Instead, it targets settings where the physical operator changes or data are limited, so preserving the model-based reconstruction structure is useful. The same teacher-guided losses could also be adapted to E2E models in future work.}
Table \ref{tab:e2e_comp} shows that while E2E methods for SPC \cite{xiang2021fista}, MRI \cite{sriram2020end}, and SR \cite{zamir2022restormer} perform well under matched test and train settings, their generalization is limited.  DIPA-PnP outperforms E2E when settings differ, demonstrating superior adaptability. DIPA-PnP only needs retraining if image resolution changes.
\begin{table}[h]
\centering
\renewcommand{\arraystretch}{1.0}
\setlength{\tabcolsep}{6pt}

\begin{tabular}{|c|c|c|c|}
\hline
\rowcolor{gray!10}
\textbf{Task} & \textbf{Testing Setting} & \textbf{E2E Model} & \textbf{L-DIPA-PnP} \\ 
\rowcolor{gray!10}
& & PSNR (dB) & PSNR (dB) \\ 
\hline
\multirow{3}{*}{\textbf{MRI}} 
& $AF_s = 3$                    & 28.23          & \textbf{32.26} \\ \cline{2-4}
& \cellcolor{blue!15} $AF_s = 5$ & \textbf{32.54} & 28.93          \\ \cline{2-4}
& $AF_s = 7$                    & 24.15          & \textbf{24.41} \\ 
\hline
\multirow{3}{*}{\textbf{CS}} 
& $\gamma_s = 0.05$                     & 17.50          & \textbf{18.11} \\ \cline{2-4}
& $\gamma_s = 0.1$                      & 21.69          & \textbf{23.80} \\ \cline{2-4}
& \cellcolor{blue!15} $\gamma_s = 0.2$  & \textbf{36.53} & 25.95          \\ 
\hline
\multirow{3}{*}{\textbf{SR}} 
& $RF_s = 1$ (Only deblurring)   & 9.25           & \textbf{28.20} \\ \cline{2-4}
& $RF_s = 2$                    & 12.07          & \textbf{20.11} \\ \cline{2-4}
& \cellcolor{blue!15} $RF_s = 4$ & \textbf{31.56} & 25.17          \\ 
\hline
\end{tabular}
\caption{Comparison with E2E models under varying testing conditions ($AF_s$, $\gamma_s$, and $RF_s$). Blue cells indicate training settings.}
\label{tab:e2e_comp}
\end{table}

\subsection{Cross-method validation} \label{app:Cross_validation}
Here, we aim to analyze the robustness of the trained PO when it is evaluated using a different method than the one on which it was originally trained.  Particularly, a PM trained with L-DIPA RED is tested with L-DIPA RED and L-DIPA PnP algorithms. We perform this evaluation for MRI, using a student with $AF_s = 5$ trained with different teachers with {$AF_t = \{1,2,3\}$} using L-DIPA RED, spatial resolution of $50 \times 50$, $\mathcal{R}_{C}(\text{\cmark})$, $\mathcal{L}_{S}(\text{\xmark})$. The results are shown in Table \ref{tab:robust}. The results show that, while the PO was trained with L-DIPA RED, using the PO with L-DIPA PnP led to a good performance (outperforming the baseline). This allows us to conclude that the PO is suitable for different algorithms. 

\begin{table}[H]
\centering
\begin{tabular}{cc|c|c|}
\hline
\multicolumn{1}{|c|}{Train\textbackslash Test}           & $AF_t$ & PnP      & RED      \\ \hline
\multicolumn{1}{|c|}{\multirow{3}{*}{RED}} & 1  & 29.65    & 27.58    \\ \cline{2-4} 
\multicolumn{1}{|c|}{}                     & 2  & 28.05    & 27.54    \\ \cline{2-4} 
\multicolumn{1}{|c|}{}                     & 3  & 29.65    & 27.31    \\ \hline
&    & Base PnP & Base RED \\
&    & 27.77    & 27.31    \\  \cline{3-4} 
\end{tabular}
\caption{Cross validation of the trained PO along different algorithms.}
\label{tab:robust}
\end{table}


\subsection{N-DIPA Linearization Analysis}
\label{app:linearization}
\begin{figure}[pos=!t]
    \centering
    \includegraphics[width=\linewidth]{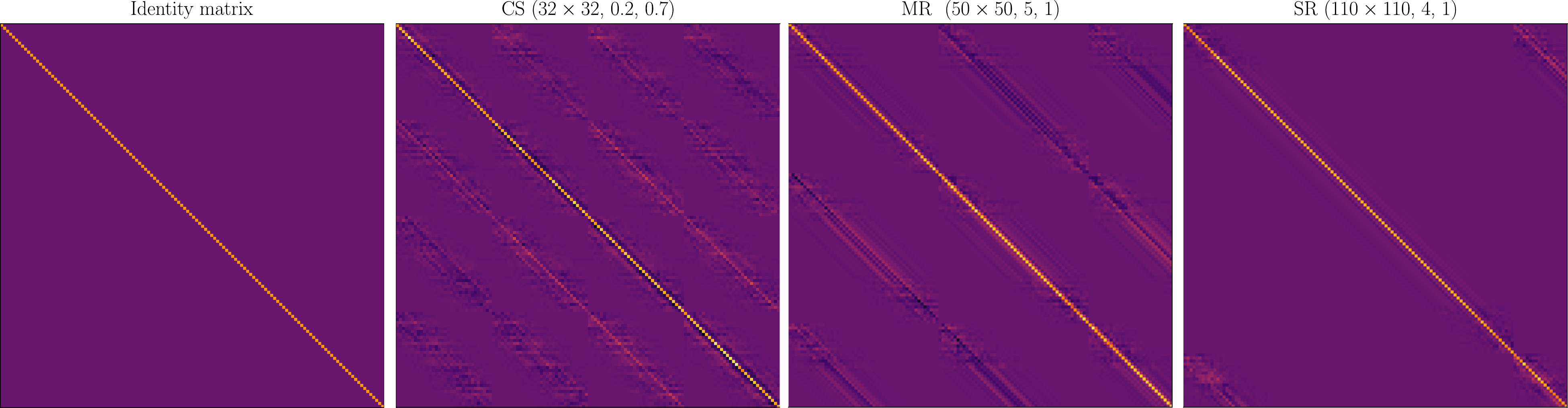}
    \vspace{-0.5cm}
    \caption{Linear representation of the learned nonlinear PO with N-DIPA.}
    \label{fig:lin}
\end{figure}

{To inspect the local structure of the nonlinear PO, we approximate its Jacobian around a reference input. This analysis is descriptive; we do not claim that the nonlinear PO is globally linear or that its eigenvalues fully characterize the algorithm.}  First, let $\mathcal{P}_{\boldsymbol{\theta}^\star} : \mathbb{R}^n \to \mathbb{R}^n$ be a differentiable, pretrained nonlinear preconditioner. In a small neighborhood around $\mathbf{x}_0$, its behavior is well-approximated by its Jacobian,
$
\left(J_{\mathcal{P}_{\boldsymbol{\theta}^\star}}(\mathbf{x}_0)\right)_{ij} = \frac{\partial (\mathcal{P}_{\boldsymbol{\theta}^\star}(\mathbf{x}_0))_i}{\partial x_j}.
$
Thus, for a small perturbation $\boldsymbol{\delta}$, we have
$
\mathcal{P}_{\boldsymbol{\theta}^\star}(\mathbf{x}_0+\boldsymbol{\delta}) \approx \mathcal{P}_{\boldsymbol{\theta}^\star}(\mathbf{x}_0) + J_{\mathcal{P}_{\boldsymbol{\theta}^\star}}(\mathbf{x}_0)\,\boldsymbol{\delta}.
$
where {$\boldsymbol{\delta} \in \mathbb{R}^n$} denotes an arbitrary small perturbation vector around $\mathbf{x}_0$. {In effect,} by injecting canonical unit vectors into the network, we obtain an approximate impulse response for each entry. Since the analytical Jacobian is typically unavailable in deep networks, we estimate it via finite differences. For a small $\epsilon>0$, the partial derivative with respect to $x_j$ is approximated as
$
\frac{\partial (\mathcal{P}_{\boldsymbol{\theta}^\star}(\mathbf{x}_0))_i}{\partial x_j} \approx \frac{(\mathcal{P}_{\boldsymbol{\theta}^\star}(\mathbf{x}_0+\epsilon\,\mathbf{e}_j))_i - (\mathcal{P}_{\boldsymbol{\theta}^\star}(\mathbf{x}_0))_i}{\epsilon},
$
where $\epsilon>0$ is a small finite-difference step size, and $\mathbf{e}_j$ is the $j$th standard basis vector in $\mathbb{R}^n$. Repeating this process for all components yields an effective approximation of the Jacobian~\cite{raghu2017expressive,dennis1996numerical}.  Based on this linearization, in Fig. \ref{fig:lin} we display the resulting matrix for each imaging task with N-DIPA, which shows similar structures to the ones of L-DIPA from Fig. \ref{fig:POs}. 

\section{{Limitations}}

One important drawback of our approach is the heavy training cost, because we must back-propagate through every iteration of the recovery loop, and optimizing the preconditioner becomes computationally intensive. {Another limitation is that DIPA cannot create information absent from the student measurements without support from the image prior learned during training. Therefore, the teacher-student transfer should be interpreted as prior-guided behavior matching on the training distribution, not as a guarantee of recovering arbitrary null-space components.} Moreover, our teacher operator \(\mathbf{A}_t\) is currently selected via heuristic rules, leveraging the known structure of the sensing matrix and empirical validation, but a principled, theory-driven criterion for choosing or learning the optimal \(\mathbf{A}_t\) remains an important open problem for future work.

\section{Conclusions}

We introduced DIPA, a teacher-guided framework for learning preconditioning operators in imaging inverse problems. Instead of designing the preconditioner solely from the physical sensing operator, DIPA uses a virtual, better-conditioned teacher operator during training to guide the reconstruction behavior of a student algorithm that uses the deployable, ill-conditioned operator. We instantiated this idea with an interpretable linear preconditioner and a scalable neural preconditioner. Experiments on MRI, single-pixel compressed sensing, and super-resolution show that DIPA improves reconstruction quality and convergence behavior relative to unpreconditioned solvers and several classical and learned preconditioning baselines. These results support the view that preconditioning can be learned not only as a numerical acceleration tool, but also as a mechanism for transferring reconstruction behavior across sensing models.

\bibliographystyle{cas-model2-names}
\bibliography{refs}

\end{document}

%% file: our_imports.tex
\usepackage{tikz,lipsum}
\usepackage[most]{tcolorbox}

\newtcolorbox{Box1}[2][]{
                lower separated=false,
                colback=white,
colframe=white!20!gray,fonttitle=\bfseries,
colbacktitle=white!10!gray,enhanced,
attach boxed title to top left={xshift=0.1cm,
        yshift=-0.01mm}, 
title=#2}
\newtheorem{theorem}{Theorem}
\makeatletter
\newcommand{\settheoremtag}[1]{
  \let\oldthetheorem\thetheorem
  \renewcommand{\thetheorem}{#1}
  \g@addto@macro\endtheorem{
    \addtocounter{theorem}{-1}
    \global\let\thetheorem\oldthetheorem}
  }
\makeatother

\newtheorem{assumption}{Assumption}
\newtheorem{example}[theorem]{Scenario}

\DeclareMathOperator*{\argmin}{arg\,min}

\usepackage{algorithm}  
\usepackage{algpseudocode}  
\usepackage{multirow}
\usepackage{rotating} 
\usepackage{makecell}
\usepackage{array}
\usepackage{tabularray}
\usepackage{pifont}
\usepackage{amssymb}
\usepackage{pifont}
\newcommand{\cmark}{\ding{51}} 
\newcommand{\xmark}{\ding{55}} 

\usepackage{colortbl}

\usepackage{graphicx}
\usepackage{amsmath}
\usepackage{booktabs}
\usepackage{tabularray}
\usepackage{ulem}
\usepackage{makecell}
\usepackage{array}
\usepackage{tabularray}
\usepackage{pifont}

%% file: refs.bib
@String(CVPR= {IEEE Conf. Comput. Vis. Pattern Recog.})

@String(ICCV= {Int. Conf. Comput. Vis.})

@String(ECCV= {Eur. Conf. Comput. Vis.})

@String(TOG= {ACM Trans. Graph.})

@String(CVPR  = {CVPR})

@String(ICCV  = {ICCV})

@String(ECCV  = {ECCV})

@String(TOG   = {ACM TOG})

@article{jin2017sparsity,
  title={Sparsity regularization in inverse problems},
  author={Jin, Bangti and Maa{\ss}, Peter and Scherzer, Otmar},
  journal={Inverse Problems},
  volume={33},
  number={6},
  year={2017},
  publisher={IOP PUBLISHING LTD}
}

@article{dassios2015preconditioner,
  title={A preconditioner for a primal-dual newton conjugate gradient method for compressed sensing problems},
  author={Dassios, Ioannis and Fountoulakis, Kimon and Gondzio, Jacek},
  journal={SIAM Journal on Scientific Computing},
  volume={37},
  number={6},
  pages={A2783--A2812},
  year={2015},
  publisher={SIAM}
}

@article{johnson1983polynomial,
  title={Polynomial preconditioners for conjugate gradient calculations},
  author={Johnson, Olin G and Micchelli, Charles A and Paul, George},
  journal={SIAM Journal on Numerical Analysis},
  volume={20},
  number={2},
  pages={362--376},
  year={1983},
  publisher={SIAM}
}

@article{duarte2008single,
  title={Single-pixel imaging via compressive sampling},
  author={Duarte, Marco F and Davenport, Mark A and Takhar, Dharmpal and Laska, Jason N and Sun, Ting and Kelly, Kevin F and Baraniuk, Richard G},
  journal={IEEE signal processing magazine},
  volume={25},
  number={2},
  pages={83--91},
  year={2008},
  publisher={IEEE}
}

@inproceedings{ronneberger2015u,
  title={U-net: Convolutional networks for biomedical image segmentation},
  author={Ronneberger, Olaf and Fischer, Philipp and Brox, Thomas},
  booktitle={Medical Image Computing and Computer-Assisted Intervention--MICCAI 2015: 18th International Conference, Munich, Germany, October 5-9, 2015, Proceedings, Part III 18},
  pages={234--241},
  year={2015},
  organization={Springer}
}

@article{golub1999tikhonov,
  title={Tikhonov regularization and total least squares},
  author={Golub, Gene H and Hansen, Per Christian and O'Leary, Dianne P},
  journal={SIAM journal on matrix analysis and applications},
  volume={21},
  number={1},
  pages={185--194},
  year={1999},
  publisher={SIAM}
}

@article{FISTA,
  title={A fast iterative shrinkage-thresholding algorithm for linear inverse problems},
  author={Beck, Amir and Teboulle, Marc},
  journal={SIAM journal on imaging sciences},
  volume={2},
  number={1},
  pages={183--202},
  year={2009},
  publisher={SIAM}
}

@INPROCEEDINGS{venkatakrishnan2013plug,
  author={Venkatakrishnan, Singanallur V. and Bouman, Charles A. and Wohlberg, Brendt},
  booktitle={2013 IEEE Global Conference on Signal and Information Processing}, 
  title={Plug-and-Play priors for model based reconstruction}, 
  year={2013},
  volume={},
  number={},
  pages={945-948},
  doi={10.1109/GlobalSIP.2013.6737048}}

@article{chan2016plug,
  title={Plug-and-play ADMM for image restoration: Fixed-point convergence and applications},
  author={Chan, Stanley H and Wang, Xiran and Elgendy, Omar A},
  journal={IEEE Transactions on Computational Imaging},
  volume={3},
  number={1},
  pages={84--98},
  year={2016},
  publisher={IEEE}
}

@inproceedings{gastal2011domain,
author = {Gastal, Eduardo S. L. and Oliveira, Manuel M.},
title = {Domain transform for edge-aware image and video processing},
year = {2011},
isbn = {9781450309431},
publisher = {Association for Computing Machinery},
address = {New York, NY, USA},
url = {https://doi.org/10.1145/1964921.1964964},
doi = {10.1145/1964921.1964964},
booktitle = {ACM SIGGRAPH 2011 Papers},
articleno = {69},
numpages = {12},
location = {Vancouver, British Columbia, Canada},
series = {SIGGRAPH '11}
}

@article{adam,
  title={Adam: A method for stochastic optimization},
  author={Kingma, Diederik P and Ba, Jimmy},
  journal={arXiv preprint arXiv:1412.6980},
  year={2014}
}

@book{bertero2021introduction,
  title={Introduction to inverse problems in imaging},
  author={Bertero, Mario and Boccacci, Patrizia and De Mol, Christine},
  year={2021},
  publisher={CRC press}
}

@article{bai2020deep,
  title={Deep learning methods for solving linear inverse problems: Research directions and paradigms},
  author={Bai, Yanna and Chen, Wei and Chen, Jie and Guo, Weisi},
  journal={Signal Processing},
  volume={177},
  pages={107729},
  year={2020},
  publisher={Elsevier}
}

@article{proximal,
  title={Proximal algorithms},
  author={Parikh, Neal and Boyd, Stephen},
  journal={Foundations and Trends in optimization},
  volume={1},
  number={3},
  pages={127--239},
  year={2014},
  publisher={Now Publishers Inc. Hanover, MA, USA}
}

@article{IPDL,
  title={Deep learning techniques for inverse problems in imaging},
  author={Ongie, Gregory and Jalal, Ajil and Baraniuk, Christopher A Metzler Richard G and Dimakis, Alexandros G and Willett, Rebecca},
  journal={IEEE Journal on Selected Areas in Information Theory},
  year={2020},
  publisher={IEEE}
}

@ARTICLE{cs,
  author={E. J. {Candes} and M. B. {Wakin}},
  journal={IEEE Signal Processing Magazine}, 
  title={An Introduction To Compressive Sampling}, 
  year={2008},
  volume={25},
  number={2},
  pages={21-30},
  doi={10.1109/MSP.2007.914731}}

@article{tian2011survey,
  title={A survey on super-resolution imaging},
  author={Tian, Jing and Ma, Kai-Kuang},
  journal={Signal, Image and Video Processing},
  volume={5},
  pages={329--342},
  year={2011},
  publisher={Springer}
}

@book{gunturk2018image,
  title={Image restoration},
  author={Gunturk, Bahadir and Li, Xin},
  year={2018},
  publisher={CRC Press}
}

@article{lustig2008compressed,
  title={Compressed sensing MRI},
  author={Lustig, Michael and Donoho, David L and Santos, Juan M and Pauly, John M},
  journal={IEEE signal processing magazine},
  volume={25},
  number={2},
  pages={72--82},
  year={2008},
  publisher={IEEE}
}

@article{zha2023learning,
  title={Learning nonlocal sparse and low-rank models for image compressive sensing: Nonlocal sparse and low-rank modeling},
  author={Zha, Zhiyuan and Wen, Bihan and Yuan, Xin and Ravishankar, Saiprasad and Zhou, Jiantao and Zhu, Ce},
  journal={IEEE Signal Processing Magazine},
  volume={40},
  number={1},
  pages={32--44},
  year={2023},
  publisher={IEEE}
}

@article{tan2024provably,
  title={Provably convergent plug-and-play quasi-Newton methods},
  author={Tan, Hong Ye and Mukherjee, Subhadip and Tang, Junqi and Sch{\"o}nlieb, Carola-Bibiane},
  journal={SIAM Journal on Imaging Sciences},
  volume={17},
  number={2},
  pages={785--819},
  year={2024},
  publisher={SIAM}
}

@article{iyer2024polynomial,
  title={Polynomial preconditioners for regularized linear inverse problems},
  author={Iyer, Siddharth S and Ong, Frank and Cao, Xiaozhi and Liao, Congyu and Daniel, Luca and Tamir, Jonathan I and Setsompop, Kawin},
  journal={SIAM Journal on Imaging Sciences},
  volume={17},
  number={1},
  pages={116--146},
  year={2024},
  publisher={SIAM}
}

@inproceedings{garber2024image,
  title={Image restoration by denoising diffusion models with iteratively preconditioned guidance},
  author={Garber, Tomer and Tirer, Tom},
  booktitle={Proceedings of the IEEE/CVF Conference on Computer Vision and Pattern Recognition},
  pages={25245--25254},
  year={2024}
}

@misc{hinton2015distillingknowledgeneuralnetwork,
      title={Distilling the Knowledge in a Neural Network}, 
      author={Geoffrey Hinton and Oriol Vinyals and Jeff Dean},
      year={2015},
      eprint={1503.02531},
      archivePrefix={arXiv},
      primaryClass={stat.ML},
      url={https://arxiv.org/abs/1503.02531}, 
}

@InProceedings{KD_SEGMENTATION,
author = {Liu, Yifan and Chen, Ke and Liu, Chris and Qin, Zengchang and Luo, Zhenbo and Wang, Jingdong},
title = {Structured Knowledge Distillation for Semantic Segmentation},
booktitle = {Proceedings of the IEEE/CVF Conference on Computer Vision and Pattern Recognition (CVPR)},
month = {June},
year = {2019}
}

@inproceedings{KD_OBJECT_DETECTION,
 author = {Chen, Guobin and Choi, Wongun and Yu, Xiang and Han, Tony and Chandraker, Manmohan},
 booktitle = {Advances in Neural Information Processing Systems},
 editor = {I. Guyon and U. Von Luxburg and S. Bengio and H. Wallach and R. Fergus and S. Vishwanathan and R. Garnett},
 pages = {},
 publisher = {Curran Associates, Inc.},
 title = {Learning Efficient Object Detection Models with Knowledge Distillation},
 url = {https://proceedings.neurips.cc/paper_files/paper/2017/file/e1e32e235eee1f970470a3a6658dfdd5-Paper.pdf},
 volume = {30},
 year = {2017}
}

@article{zagoruyko2016paying,
  title={Paying more attention to attention: Improving the performance of convolutional neural networks via attention transfer},
  author={Zagoruyko, Sergey and Komodakis, Nikos},
  journal={arXiv preprint arXiv:1612.03928},
  year={2016}
}

@article{bertero1985linear,
  title={Linear inverse problems with discrete data. I. General formulation and singular system analysis},
  author={Bertero, Mario and De Mol, Christine and Pike, Edward Roy},
  journal={Inverse problems},
  volume={1},
  number={4},
  pages={301},
  year={1985},
  publisher={IOP Publishing}
}

@article{zhang2021plug,
  title={Plug-and-play image restoration with deep denoiser prior},
  author={Zhang, Kai and Li, Yawei and Zuo, Wangmeng and Zhang, Lei and Van Gool, Luc and Timofte, Radu},
  journal={IEEE Transactions on Pattern Analysis and Machine Intelligence},
  volume={44},
  number={10},
  pages={6360--6376},
  year={2021},
  publisher={IEEE}
}

@inproceedings{ma2023improving,
  title={Improving Medical Image Denoising via a Lightweight Plug-and-play Module},
  author={Ma, Lei and Kuang, Hulin and Liu, Jin and Shen, Chengchao and Wang, Jianxin},
  booktitle={2023 IEEE International Conference on Bioinformatics and Biomedicine (BIBM)},
  pages={1350--1355},
  year={2023},
  organization={IEEE}
}

@article{teodoro2019image,
  title={Image restoration and reconstruction using targeted plug-and-play priors},
  author={Teodoro, Afonso M and Bioucas-Dias, Jos{\'e} M and Figueiredo, M{\'a}rio AT},
  journal={IEEE Transactions on Computational Imaging},
  volume={5},
  number={4},
  pages={675--686},
  year={2019},
  publisher={IEEE}
}

@inproceedings{BM3D,
  title={Image denoising with block-matching and 3D filtering},
  author={Dabov, Kostadin and Foi, Alessandro and Katkovnik, Vladimir and Egiazarian, Karen},
  booktitle={Image processing: algorithms and systems, neural networks, and machine learning},
  volume={6064},
  pages={354--365},
  year={2006},
  organization={SPIE}
}

@article{bcm_nlm,
    title   = {{Non-Local Means Denoising}},
    author  = {Buades, Antoni and Coll, Bartomeu and Morel, Jean-Michel},
    journal = {{Image Processing On Line}},
    volume  = {1},
    pages   = {208--212},
    year    = {2011}
}

@inproceedings{burger2012image,
  title={Image denoising: Can plain neural networks compete with BM3D?},
  author={Burger, Harold C and Schuler, Christian J and Harmeling, Stefan},
  booktitle={2012 IEEE conference on computer vision and pattern recognition},
  pages={2392--2399},
  year={2012},
  organization={IEEE}
}

@article{boyd2011distributed,
  title={Distributed optimization and statistical learning via the alternating direction method of multipliers},
  author={Boyd, Stephen and Parikh, Neal and Chu, Eric and Peleato, Borja and Eckstein, Jonathan and others},
  journal={Foundations and Trends{\textregistered} in Machine learning},
  volume={3},
  number={1},
  pages={1--122},
  year={2011},
  publisher={Now Publishers, Inc.}
}

@article{he2013half,
  title={Half-quadratic-based iterative minimization for robust sparse representation},
  author={He, Ran and Zheng, Wei-Shi and Tan, Tieniu and Sun, Zhenan},
  journal={IEEE transactions on pattern analysis and machine intelligence},
  volume={36},
  number={2},
  pages={261--275},
  year={2013},
  publisher={IEEE}
}

@article{daubechies2004iterative,
  title={An iterative thresholding algorithm for linear inverse problems with a sparsity constraint},
  author={Daubechies, Ingrid and Defrise, Michel and De Mol, Christine},
  journal={Communications on Pure and Applied Mathematics: A Journal Issued by the Courant Institute of Mathematical Sciences},
  volume={57},
  number={11},
  pages={1413--1457},
  year={2004},
  publisher={Wiley Online Library}
}

@article{gou2021knowledge,
  title={Knowledge distillation: A survey},
  author={Gou, Jianping and Yu, Baosheng and Maybank, Stephen J and Tao, Dacheng},
  journal={International Journal of Computer Vision},
  volume={129},
  number={6},
  pages={1789--1819},
  year={2021},
  publisher={Springer}
}

@article{strong2003edge,
  title={Edge-preserving and scale-dependent properties of total variation regularization},
  author={Strong, David and Chan, Tony},
  journal={Inverse problems},
  volume={19},
  number={6},
  pages={S165},
  year={2003},
  publisher={IOP Publishing}
}

@article{RED,
  title={The little engine that could: Regularization by denoising (RED)},
  author={Romano, Yaniv and Elad, Michael and Milanfar, Peyman},
  journal={SIAM Journal on Imaging Sciences},
  volume={10},
  number={4},
  pages={1804--1844},
  year={2017},
  publisher={SIAM}
}

@article{dahlke2012multilevel,
  title={Multilevel preconditioning and adaptive sparse solution of inverse problems},
  author={Dahlke, Stephan and Fornasier, Massimo and Raasch, Thorsten},
  journal={Mathematics of Computation},
  volume={81},
  number={277},
  pages={419--446},
  year={2012}
}

@article{nielsen2010efficient,
  title={Efficient preconditioners for optimality systems arising in connection with inverse problems},
  author={Nielsen, Bj{\o}rn Fredrik and Mardal, Kent-Andre},
  journal={SIAM Journal on Control and Optimization},
  volume={48},
  number={8},
  pages={5143--5177},
  year={2010},
  publisher={SIAM}
}

@inproceedings{murugesan2020kd,
  title={KD-MRI: A knowledge distillation framework for image reconstruction and image restoration in MRI workflow},
  author={Murugesan, Balamurali and Vijayarangan, Sricharan and Sarveswaran, Kaushik and Ram, Keerthi and Sivaprakasam, Mohanasankar},
  booktitle={Medical imaging with deep learning},
  pages={515--526},
  year={2020},
  organization={PMLR}
}

@inproceedings{li2022multiple,
  title={Multiple degradation and reconstruction network for single image denoising via knowledge distillation},
  author={Li, Juncheng and Yang, Hanhui and Yi, Qiaosi and Fang, Faming and Gao, Guangwei and Zeng, Tieyong and Zhang, Guixu},
  booktitle={Proceedings of the IEEE/CVF conference on computer vision and pattern recognition},
  pages={558--567},
  year={2022}
}

@inproceedings{pruessmann1998coil,
  title={Coil sensitivity encoding for fast MRI},
  author={Pruessmann, Klaas P and Weiger, Markus and Scheidegger, Markus B and Boesiger, Peter},
  booktitle={Proceedings of the ISMRM 6th Annual Meeting, Sydney},
  volume={1998},
  year={1998}
}

@article{loshchilov2017decoupled,
  title={Decoupled weight decay regularization},
  author={Loshchilov, I},
  journal={arXiv preprint arXiv:1711.05101},
  year={2017}
}

@article{chen2021multiscale,
  title={Multiscale Cholesky preconditioning for ill-conditioned problems},
  author={Chen, Jiong and Sch{\"a}fer, Florian and Huang, Jin and Desbrun, Mathieu},
  journal={ACM Transactions on Graphics (TOG)},
  volume={40},
  number={4},
  pages={1--13},
  year={2021},
  publisher={ACM New York, NY, USA}
}

@article{anzt2019adaptive,
  title={Adaptive precision in block-Jacobi preconditioning for iterative sparse linear system solvers},
  author={Anzt, Hartwig and Dongarra, Jack and Flegar, Goran and Higham, Nicholas J and Quintana-Ort{\'\i}, Enrique S},
  journal={Concurrency and Computation: Practice and Experience},
  volume={31},
  number={6},
  pages={e4460},
  year={2019},
  publisher={Wiley Online Library}
}

@inproceedings{gander2017origins,
  title={On the origins of linear and non-linear preconditioning},
  author={Gander, Martin J},
  booktitle={Domain decomposition methods in science and engineering XXIII},
  pages={153--161},
  year={2017},
  organization={Springer}
}

@article{doi:10.1137/17M1128502,
author = {Liu, Lulu and Keyes, David E. and Krause, Rolf},
title = {A Note on Adaptive Nonlinear Preconditioning Techniques},
journal = {SIAM Journal on Scientific Computing},
volume = {40},
number = {2},
pages = {A1171-A1186},
year = {2018},
doi = {10.1137/17M1128502},

URL = { 
    
        https://doi.org/10.1137/17M1128502
    
    

},
eprint = { 
    
        https://doi.org/10.1137/17M1128502
    
    

}
}

@inproceedings{andrea2016preconditioning,
  title={Preconditioning strategies for nonlinear conjugate gradient methods, based on quasi-Newton updates},
  author={Andrea, Caliciotti and Giovanni, Fasano and Massimo, Roma},
  booktitle={AIP conference proceedings},
  volume={1776},
  number={1},
  year={2016},
  organization={AIP Publishing}
}

@article{pearson2020preconditioners,
  title={Preconditioners for Krylov subspace methods: An overview},
  author={Pearson, John W and Pestana, Jennifer},
  journal={GAMM-Mitteilungen},
  volume={43},
  number={4},
  pages={e202000015},
  year={2020},
  publisher={Wiley Online Library}
}

@article{chan1984nonlinearly,
  title={Nonlinearly preconditioned Krylov subspace methods for discrete Newton algorithms},
  author={Chan, Tony F and Jackson, Kenneth R},
  journal={SIAM Journal on scientific and statistical computing},
  volume={5},
  number={3},
  pages={533--542},
  year={1984},
  publisher={SIAM}
}

@article{fessler1999conjugate,
  title={Conjugate-gradient preconditioning methods for shift-variant PET image reconstruction},
  author={Fessler, Jeffrey A and Booth, Scott D},
  journal={IEEE transactions on image processing},
  volume={8},
  number={5},
  pages={688--699},
  year={1999},
  publisher={IEEE}
}

@ARTICLE{FastMRI_dataset,
  title     = "{FastMRI}: A publicly available raw k-space and {DICOM} dataset
               of knee images for accelerated {MR} image reconstruction using
               machine learning",
  author    = "Knoll, Florian and Zbontar, Jure and Sriram, Anuroop and
               Muckley, Matthew J and Bruno, Mary and Defazio, Aaron and
               Parente, Marc and Geras, Krzysztof J and Katsnelson, Joe and
               Chandarana, Hersh and Zhang, Zizhao and Drozdzalv, Michal and
               Romero, Adriana and Rabbat, Michael and Vincent, Pascal and
               Pinkerton, James and Wang, Duo and Yakubova, Nafissa and Owens,
               Erich and Zitnick, C Lawrence and Recht, Michael P and
               Sodickson, Daniel K and Lui, Yvonne W",
  journal   = "Radiol. Artif. Intell.",
  publisher = "Radiological Society of North America (RSNA)",
  volume    =  2,
  number    =  1,
  pages     = "e190007",
  month     =  jan,
  year      =  2020,
  language  = "en"
}

@article{deng2012mnist,
  title={The mnist database of handwritten digit images for machine learning research [best of the web]},
  author={Deng, Li},
  journal={IEEE signal processing magazine},
  volume={29},
  number={6},
  pages={141--142},
  year={2012},
  publisher={IEEE}
}

@misc{Tachella_DeepInverse_A_deep_2023,
author = {Tachella, Julian and Chen, Dongdong and Hurault, Samuel and Terris, Matthieu and Wang, Andrew},
doi = {10.5281/zenodo.7982256},
month = jun,
title = {{DeepInverse: A deep learning framework for inverse problems in imaging}},
url = {https://github.com/deepinv/deepinv},
version = {latest},
year = {2023}
}

@inproceedings{liu2015faceattributes,
  title = {Deep Learning Face Attributes in the Wild},
  author = {Liu, Ziwei and Luo, Ping and Wang, Xiaogang and Tang, Xiaoou},
  booktitle = {Proceedings of International Conference on Computer Vision (ICCV)},
  month = {December},
  year = {2015} 
}

@article{arbelaez2010contour,
  title={Contour detection and hierarchical image segmentation},
  author={Arbelaez, Pablo and Maire, Michael and Fowlkes, Charless and Malik, Jitendra},
  journal={IEEE transactions on pattern analysis and machine intelligence},
  volume={33},
  number={5},
  pages={898--916},
  year={2010},
  publisher={IEEE}
}

@inproceedings{zamir2022restormer,
  title={Restormer: Efficient transformer for high-resolution image restoration},
  author={Zamir and others},
  booktitle={IEEE/CVF CVPR},
  pages={5728--5739},
  year={2022}
}

@article{xiang2021fista,
  title={FISTA-Net: Learning a fast iterative shrinkage thresholding network for inverse problems in imaging},
  author={Xiang and others},
  journal={IEEE Trans. Med. Imaging},
  volume={40},
  number={5},
  pages={1329--1339},
  year={2021},
  publisher={IEEE}
}

@article{kamilov2023plug,
  title={Plug-and-play methods for integrating physical and learned models in computational imaging: Theory, algorithms, and applications},
  author={Kamilov and others},
  journal={IEEE Sig. Proc. Mag.},
  volume={40},
  number={1},
  pages={85--97},
  year={2023},
  publisher={IEEE}
}

@inproceedings{sriram2020end,
  title={End-to-end variational networks for accelerated MRI reconstruction},
  author={Sriram and others},
  booktitle={MICCAI 2020},
  pages={64--73},
  year={2020},
  organization={Springer}
}

@article{ehrhardt2024learning,
  title={Learning preconditioners for inverse problems},
  author={Ehrhardt and others},
  journal={arXiv preprint arXiv:2406.00260},
  year={2024}
}

@article{rosenblatt1958perceptron,
  title={The perceptron: a probabilistic model for information storage and organization in the brain.},
  author={Rosenblatt, Frank},
  journal={Psychological review},
  volume={65},
  number={6},
  pages={386},
  year={1958},
  publisher={American Psychological Association}
}

@article{lecun1998gradient,
  title={Gradient-based learning applied to document recognition},
  author={LeCun, Yann and Bottou, L{\'e}on and Bengio, Yoshua and Haffner, Patrick},
  journal={Proceedings of the IEEE},
  volume={86},
  number={11},
  pages={2278--2324},
  year={1998},
  publisher={Ieee}
}

@inproceedings{woo2018cbam,
  title={Cbam: Convolutional block attention module},
  author={Woo, Sanghyun and others},
  booktitle={Proceedings of the European conference on computer vision (ECCV)},
  pages={3--19},
  year={2018}
}

@article{yuan2018multiscale,
  title={A multiscale and multidepth convolutional neural network for remote sensing imagery pan-sharpening},
  author={Yuan, Qiangqiang and others},
  journal={IEEE Journal of Selected Topics in Applied Earth Observations and Remote Sensing},
  volume={11},
  number={3},
  pages={978--989},
  year={2018},
  publisher={IEEE}
}

@article{delbracio2023inversion,
  title={Inversion by direct iteration: An alternative to denoising diffusion for image restoration},
  author={Delbracio, Mauricio and Milanfar, Peyman},
  journal={arXiv preprint arXiv:2303.11435},
  year={2023}
}

@article{dosovitskiy2020image,
  title={An image is worth 16x16 words: Transformers for image recognition at scale},
  author={Dosovitskiy, Alexey and others},
  journal={arXiv preprint arXiv:2010.11929},
  year={2020}
}

@inproceedings{liu2022convnet,
  title={A convnet for the 2020s},
  author={Liu, Zhuang and others},
  booktitle={Proceedings of the IEEE/CVF conference on computer vision and pattern recognition},
  pages={11976--11986},
  year={2022}
}

@inproceedings{raghu2017expressive,
  title={On the expressive power of deep neural networks},
  author={Raghu, Maithra and others},
  booktitle={international conference on machine learning},
  pages={2847--2854},
  year={2017},
  organization={PMLR}
}

@book{dennis1996numerical,
  title={Numerical methods for unconstrained optimization and nonlinear equations},
  author={Dennis Jr, John E and Schnabel, Robert B},
  year={1996},
  publisher={SIAM}
}

@article{jacome2026npn,
  title={{NPN: Non-Linear Projections of the Null-Space for Imaging Inverse Problems}},
  author={Jacome, Roman and Gualdr{\'o}n-Hurtado, Romario and Su{\'a}rez-Rodr{\'\i}guez, Le{\'o}n and Arguello, Henry},
  journal={Advances in Neural Information Processing Systems},
  volume={38},
  pages={119069--119099},
  year={2026}
}

@article{gualdron2026gsnr,
  title={{GSNR: Graph Smooth Null-Space Representation for Inverse Problems}},
  author={Gualdr{\'o}n-Hurtado, Romario and Jacome, Roman and Suarez, Rafael S and Arguello, Henry},
  journal={arXiv preprint arXiv:2602.20328},
  year={2026}
}

@INPROCEEDINGS{gualdron2025deep, author={Gualdrón-Hurtado, Romario and Jacome, Roman and Suarez, Leon and Galvis, Laura and Arguello, Henry}, booktitle={2025 IEEE 10th International Workshop on Computational Advances in Multi-Sensor Adaptive Processing (CAMSAP)}, title={Deep Distillation Gradient Preconditioning for Inverse Problems}, year={2025}, volume={}, number={}, pages={166-170}, doi={10.1109/CAMSAP66162.2025.11423951}}
